\documentclass[aps,prd,showpacs,nofootinbib,preprint,tightenlines]{revtex4}
\DeclareFontFamily{U}{rsfs}{}         
\DeclareFontShape{U}{rsfs}{m}{n}{<5> rsfs5 <6><7> rsfs7          %
  <8><9><10><10.95><12><14.4><17.28><20.74><24.88> rsfs10}{}     %
\DeclareMathAlphabet{\mathfs}{U}{rsfs}{m}{n}                     %
\newcommand{\mfs}[1]{\mathfs {#1}}                               %

\usepackage{amsmath,bbm}
\usepackage{amssymb}
\newcommand{\va}{\scriptscriptstyle}

\newcommand{\sH}{{\mfs H}}
\newcommand{\sL}{{\mfs L}}

\def\be{\nopagebreak[3]\begin{equation}}
\def\ee{\end{equation}}
\def\ba{\nopagebreak[3]\begin{eqnarray}}
\def\ea{\end{eqnarray}}

\def\l{\langle}
\def\r{\rangle}

\def\H{{\cal H}}

\newcommand{\C}{\mathbb{C}}

\newcommand{\R}{\mathbb{R}}
\newcommand{\Z}{\mathbb{Z}}
\newcommand{\teta}{\rlap{\lower2ex\hbox{$\,\tilde{}$}}\eta{}}

\newcounter{mnotecount}[section]

\begin{document}
\preprint{\vbox{\baselineskip=12pt \rightline{ICN-UNAM-05/01}
\rightline{gr-qc/yymmnnn} }}
\title{ On the symmetry of the vacuum in theories with spontaneous symmetry breaking}


\author{Alejandro Perez$^{1}$\thanks{perez@gravity.psu.edu},
  and Daniel Sudarsky$^{2}$\thanks{sudarsky@nuclecu.unam.mx}\\[.5cm]
  1. Centre de Physique Th\'eorique (Unit\'e Mixte de Recherche (UMR 6207) du CNRS et des
Universit\'es Aix-Marseille I, Aix-Marseille II, et du Sud
Toulon-Var; laboratoire afili\'e \`a la FRUMAM (FR 2291)), Campus de
Luminy, 13288 Marseille,
France. \\
  2. Instituto de Ciencias Nucleares\\
  Universidad Nacional Aut\'onoma de M\'exico\\
  A. Postal 70-543, M\'exico D.F. 04510, M\'exico\\
}

\begin{abstract}
We review the usual account of the phenomena of spontaneous
symmetry breaking (SSB), pointing out the common misunderstandings
surrounding the issue,  in particular within the context of
quantum field theory. In fact, the common explanations one  finds
in this context, indicate that   under  certain  conditions
corresponding to the situation called  SSB, the vacuum of the
theory does not share the symmetries of the Lagrangian. We explain
in detail why this statement is incorrect in general, and in what
limited  set of  circumstances such situation  could arise.   We
concentrate on the case of global symmetries, for which  we found
no satisfactory exposition in   the existing literature, and
briefly comment on the  case of gauge symmetries where, although
insufficiently publicized,  accurate  and  complete descriptions
exist. We briefly discuss the implications for the
phenomenological manifestations usually attributed to the
phenomena of spontaneous symmetry breaking, analyzing which might
be affected by our analysis  and which are not. In particular we
describe the mass generation mechanism in a fully symmetric scheme
(i.e.,  with a  totally symmetric vacuum), and briefly discuss the
implications of this analysis to the  problem  of formation of
topological defects in the early universe.
\end{abstract}

 \maketitle

\section{Introduction}
The understanding of theories with ``spontaneous symmetry
breaking" (SSB) is currently taken as belonging to what one might
call ``the established part" of theoretical physics,  not less due
to the fact that it comprises nowadays  a fundamental aspect of
the standard model of particle physics, and plays  an important
role in the understanding of well known phenomena
 like super-fluidity,  superconductivity,  and of the behavior of   multitude of  systems  displaying phase
 transitions of various kinds. In fact a Nobel price  was just  awarded to 
 work on the subject.  It is thus surprising that something is
left to be said  on this subject at this point in time. In fact,
it well might be that the points we  will be  making  in this
manuscript are well understood by some of the experts on the
field, nevertheless we find it remarkable that no fully
satisfactory  exposition of the subject can be found in the
literature, and that most people working in related areas do not
seem to have a complete  understanding of the issue at hand.

 Before  we take on this discussion it its convenient to reassure the reader that 
 { \bf all the successful phenomenology usually attributed to SSB  will be recovered in the picture we
  will present here}.  The  well established physical conclusions will be unaltered despite the fact that we 
  will argue that the standard picture is  not only confusing but also misleading 
  in various aspects, and that all the standard results  can be understood 
   within the context  of   the analysis we will be presenting throughout this manuscript.

In  almost all presentations of the subject one is informed that
one is dealing with a situation in which the symmetries of the
theory are not shared by the ground state of the system, the
standard example being of course a system with one degree of
freedom, standard kinetic term and double well potential:
classically, the lowest energy states correspond to the system at
rest  at one of the two minima of the potential. The problem
arises when we want to treat the system quantum mechanically, as
it is well known, the ground state does not correspond to a
localized wave function  at either of the two minima but rather to
a wave function representing a symmetric superposition of
localized wave packets. In this (finitely many degrees of freedom)
example, the quantum ground state does indeed share the symmetries
of the theory. When extrapolating this result to the field theory
case, the questions one is forced to confront are: does the
analogy with the one degree of freedom work?,  if not, why not?
and if it does, what happens then to all the phenomenology one
learns to associate with the existence of the large set of
vacuua---all of them asymmetric, among which, a single
one\footnote{In fact the issue of which one is not deem to be
relevant as they are all equivalent in the sense of being
 connected amongst themselves  by symmetry transformations.} is  thought to represent the state
of our universe (or region thereof)---such as Goldstone bosons,
the emergence of masses for particles in the standard model, and
the prediction of topological defects resulting from the phase
transitions in the early universe?

We will see, in this paper, that most of the standard predictions
usually associated with SSB remain unchanged when these issues are
dealt with, taking into account the quantum nature of the vacua, and
its resulting symmetries. That is, we will argue that in many
situations usually associated with a ``spontaneous breaking of the
symmetry", the  {\bf true vacuum is in fact symmetric}, and that this modifies in no
way most of the standard results usually associated with SSB.
One of the few exceptions seems to be the prediction of topological
defects, as inevitable result of the breaking of symmetries in the
early universe.  In fact as it is well known, no such objects have
ever been observed in association with the fundamental theories of
particle physics, but only in connection with the phase transitions in
systems described with statistical mechanics. We will see that, in a
certain sense, the last two are fundamentally different and will
explain how those differences might account for the emergence of
topological defects in one case and not in the other.

The paper is organized as follows. In the reminder of this  section we consider the issue of symmetries of the ground state in quantum mechanical systems.  In section II  we consider the issue in the   field theoretical case of global symmetries.  In section III  we give a review of the appropriate treatment of the problem in the case of gauge symmetries. In section IV   we consider  the problem of spontaneous symmetry breaking in  statistical mechanics. In section V   we examine  the relevance of the issues treated here  to the topic of  of topological defects,  and we end  with a brief discussion in section VI.

 \subsection{Symmetry breaking in isolated quantum mechanical
 systems}\label{QM}

Let us discuss the question of symmetry breaking in the context of
standard quantum mechanics.  We start by considering the case of a
system with two degrees of  freedom possessing a $U(1)$ symmetry:
for instance a two
   dimensional anharmonic oscillator with Lagrangian
   \be L= \frac{1}{2} m (
   \dot X_1^2 +\dot X_2^2 ) -\lambda ( X_1^2 +\ X_2^2 -v^2)^2
    \label{unos}
    \ee
    In
   order to find its ground state it is useful to change to polar
   variables $r, \theta$ such that $X_1 = r \sin(\theta)$ and $X_2 = r
   \cos(\theta)$, where as usual, we must recall the limitations of this
   change of coordinates reflected in the fact that we must keep the
   identification of $\theta$  and   $\theta + 2\pi$.
The minimum of the potential corresponds  to the circle $X_1^2 +\
X_2^2=v$. Therefore, classically there is a circle-worth of ground
states related by the $U(1)$ symmetry transformation. Classically,
the possible ground states of the system break in this sense the
$U(1)$ symmetry of the Hamiltonian. When this happens we will say
that a  classical spontaneous symmetry breaking (CSSB) is at play.

Let us now go to the quantum theory.  The Lagrangian for the
system is now written as 
\be
   \label{dos}
 L= \frac{1}{2} m ( \dot
  r^2 + r^2 \dot \theta^2 ) -\lambda ( r^2 -v^2)^2, 
  \ee 
  and the Hamiltonian
  is 
  \be
  \label{three} H= \frac{1}{2m}\ P_r^2 + \frac{1}{2mr^2}P_ \theta^2 +\lambda
  ( r^2 -v^2)^2.
   \ee 
   Upon quantization the
  ground state of the system is:
 \be \Psi_0 (r, \theta)= \frac{1}{\sqrt{2\pi}}\Phi_0(r),
   \label{wf0}
    \ee where $\Phi_0(r)$ is the ground state wave function for the
 one degree of freedom system with Hamiltonian $ H^{(1)}
 =\frac{1}{2m}\ P_r^2 +\lambda ( r^2 -v^2)^2$.  In other words, the
 ground state wave function $ \Psi_0 (r, \theta)$ is evenly
 distributed among the values of $\theta$ and is thus invariant under
 rotations as is the Hamiltonian. It is easy to see that any
 localization of the variable $\theta$, would lead to an increase in
 the expectation value of the energy of the system.

 If $\lambda$ is large enough so that $\Phi_0(r) $, is sharply peaked
  around $r=v$, then we can write some  eigenstates
  of the Hamiltonian, which are close to the vacuum,  approximately  as $ \Psi (r, \theta)= \frac{1}{2\pi}\Phi_0(r)
  e^{i k \theta } $ were $ k = n/2 \pi$, with $n$ integer. If $E_0$ is
  the energy of the ground state, then these states have approximate
  energies given by $ E_0 +\frac{k^2}{2mv^2}$.  This degree of freedom
  has only a kinetic term and thus would correspond to what we
  normally would call a ``Goldstone Boson", in a field theory
  analoge. In fact we can write the approximate effective
  Hamiltonian
  for this angular degree of freedom as
  \be
   H_{eff} (\theta,
  P_\theta)= \langle \Phi_0| H | \Phi_0\rangle = \langle \Phi_0 |
  H^{(1)} | \Phi_0\rangle + \langle \Phi_0 |\frac{1}{2mr^2}P_ \theta^2
  |\Phi_0 \rangle \approx E_0 +\frac{1}{2mv^2}P_ \theta^2. 
   \label{cinco}
  \ee
Of course the fact that the degree of freedom $\theta$ is free is
already apparent from the exact Hamiltonian (\ref{three}). The
language of effective Hamiltonian will be more useful in the
following example.

Now let's consider a system of three degrees of freedom to
illustrate the 
 mass generating mechanism associated with
``spontaneous symmetry breaking". The Lagrangian is \be L'= \frac{1}{2} m
( \dot X_1^2 +\dot X_2^2 ) + (\mu/2)\dot X_3^2 -\lambda ( X_1^2 +\
X_2^2 -v^2)^2 - \alpha\ (X_1^2 +\ X_2^2) X_3^2 \ee Note that there
is no term that is simply quadratic in $X_3$ so this degree of
freedom can not be though of as associated with an harmonic
oscillator. The Hamiltonian  (after the same change of variables
as before)  can be written as:
 \be H'= \frac{1}{2m}\ P_r^2 + \frac{1}{2mr^2}P_
\theta^2 + \frac{1}{2\mu}P_3^2 +\lambda ( r^2 -v^2)^2 +\alpha r^2
X_3^2. \ee
Now the wave function of the ground state of the system
can be written as
 \be
  \Psi'_0 (r, \theta, X_3)= \frac{1}{2\pi}\Phi'_0(r, X_3),
   \ee
where $\Phi_0' (r, X_3)$ is the ground state wave function for the
two degree of freedom system with Hamiltonian $ H^{(2)}
=\frac{1}{2m} {P_r}^2 +\lambda ( r^2 -v^2)^2 + \frac{1}{2\mu}P_3^2
+\alpha r^2 X_3^2$.  Again the wave function is independent of $
\theta$ and thus invariant under rotations in the $X_1- X_2$
plane.  We might use the fact that $ H'(r,\theta,X_3, P_r,
P_\theta, P_3) = H(r,\theta, P_r, P_\theta, ) +
\frac{1}{2\mu}P_3^2 +\alpha r^2 X_3^2$ to write an approximate
effective Hamiltonian for the degree of freedom $X_3$ in the
ground state (of the other variables) as $ H'_{eff} (X_3, P_3)=
\langle \Psi_0| H' |\Psi_0\rangle $ where $\Psi_0 $ is the ground
state wave function of eq. (\ref{wf0}), thus 
\be H'_{eff} (X_3,
P_3)= \langle \Psi_0| H |\Psi_0\rangle + \langle \Psi_0 |
(\frac{1}{2\mu}P_3^2 +\alpha r^2 X_3^2)| \Psi_0 \rangle \approx
E_0 + \frac{1}{2\mu}P_3^2 +\alpha v^2 X_3^2 \ee
 which now exhibits
the standard harmonic oscillator form, with spring constant $K= 2
\alpha v^2$.  We have therefore given a non vanishing spring
constant (the analog of mass in field theoretical Higgs mechanism)
to $X_3$ without breaking the symmetry. This simple quantum
mechanical example contains the main idea  whose extension  to
field theory  will be considered in the remainder of this
manuscript.  In fact we  will argue in the following  that in most
of the cases considered paradigmatic of the phenomena of SSB, the
physical vacuum, is indeed symmetric.

We just illustrated how continuous symmetries are unbroken by the
ground state in standard quantum
mechanics, for  a large class of  systems exhibiting CSSB. At this point we want to consider
the question: If symmetry is
not broken by the ground state in QM, is there a characteristic
feature of the ground state in systems with CSSB? The answer
should be apparent:

\vskip.5cm{\em{\bf Observation 1:} In systems with finitely many
degrees of freedom exhibiting CSSB the quantum ground state is
symmetric, and therefore, highly non-semiclassical.} \vskip.5cm

 Roughly speaking, the ground state cannot be picked
around some classical configuration as it must give equal
amplitude to an infinite set of classical configurations associate
to the classical ground states of the system. In particular in the
example of this section is  clear that the ground state wave
function (\ref{wf0}) is very far from  satisfying any notion of
semi-classicality.
In particular the semiclassical property
\be
\langle\psi | X_1^2 +\ X_2^2 |\psi \rangle=
\langle\psi|X_1 |\psi
\rangle\langle\psi|X_1 |\psi \rangle +
\langle\psi|X_2 |\psi
\rangle\langle\psi|X_2 |\psi \rangle
+O(\hbar)\ee is highly violated
for the ground state (\ref{wf0}), as  we have $\langle\psi|
X_1^2 +\ X_2^2 |\psi\rangle\approx v^2$ while $\langle\psi|X_1 |\psi\rangle=\langle\psi|X_2 |\psi\rangle=0$.
This  is a key feature of the quantum ground state of systems with CSSB: it implies that we cannot investigate the properties of the
vacuum state of such systems using semiclassical analysis.
Nevertheless, there are text books where the analog of the
previous condition in field theory is used to argue for the non
vanishing of the vacuum expectation value of the analog of $X_2$ (see eq. \ref{noway}); a
signal of SSB \cite{Nair} eq. 12.9. The point is that the ground
state of a quantum system is a very quantum mechanical state, the
assignment  to it, of semiclassical properties is very delicate  issue in
general, and  is mostly wrong in the context of systems with CSSB in
particular.

Before going into the field theoretical context let us illustrate
another important point with a different toy model. Consider a
system with Hamiltonian $H$ defined on a two dimensional Hilbert
space ${\sH}$ with a discrete symmetry represented by a unitary
operator $\hat P :{\sH} \to{\sH}$ with $ PHP^{\dagger} =H$. This
toy model corresponds to a minimalistic version of the double well
potential problem mentioned in the introduction. We choose a basis
of two orthonormal states $|R\rangle$ and $|L\rangle$
Let us
further assume that $H$ is symmetric under the replacement
$|R\rangle \leftrightarrow |L\rangle$. By symmetry, one would then
expect the ground state to be some linear combination of these two
states with equal probabilities: $ |0\rangle =2^{-1/2} (|R\rangle
+ e^{i\alpha}|L\rangle)$. The issue is what would choose the value
of the phase $\alpha$? The answer lies in the Hamiltonian. Suppose
$\langle R|H|L\rangle = c\not=0$. Let us write $c=|c|e^{i\beta}$
Consider then the value of $\langle0|H|0\rangle = (1/2)[\langle
R|H|R\rangle + \langle L|H|L\rangle + 2 Re ( e^{i\alpha}\langle
R|H|L\rangle )]$ the assumption that H is symmetric under the
replacement $|R\rangle \leftrightarrow |L\rangle$ implies that
$\langle R|H|R\rangle = \langle L|H|L\rangle =h$ where $h$ is
 a real number.  Then we have
$\langle0|H|0\rangle = h + |c|\cos (\alpha +\beta)$ .  So the
requirement that the ground state should have the minimum
expectation value for the energy leads to a unique choice for
$\alpha$ namely $\alpha = \pi - \beta$.

What happens if $c=0$? In that case the vacuum becomes degenerate,
as the value of $\alpha$ is undetermined. In addition  the  non
symmetric states  $|R\rangle$ or $|L\rangle$ are also possible ground
states. This situation corresponds then to the case where, if the
system was prepared initially in the state $|R\rangle $ it would
have zero probability of ``tunneling" to the state $|L\rangle$.
However, if the value of $c$ is determined by some physical
mechanism, for which $c=0$ represents some idealized situation
(for example, we can think of the limiting situation where the
central bump of the double well potential becomes of infinite
hight adiabatically in an example  with  time dependent Hamiltonian), then the degeneracy of the vacuum at $c=0$
can be understood as a singular feature of that limit. In other
words, no matter how small $c$ is the vacumm is uniquely
determined and induces a unique choice of  vacuum in the
limit \footnote{ We must be careful not to confuse the limit $c\rightarrow 0$  with the case $c=0$ as there is in general no warrantee of ``continuity".}$c\rightarrow 0$,  which is, in fact,  symmetric. If $c=0$ does not correspond to such
idealized situation (for instance if there is a barrier  which is actually of  infinite hight), then the degeneracy is genuinely there.
Nevertheless, among the set of ground states we still have
 the symmetric states (the only exception seems to  corresponds  situations where the Hilbert space  itself contains no symmetric states at all, as discussed in section V.B ). In the cases  where there are degenerate ground states, one could think of evoking
super-selection rules\footnote{ \label{ssrules} We must  be careful with the usage of the  words ``super-selection rules"  as their original meaning was to  be attached to situations where the  superposition  principle  was to  restrict the application of the  superposition principle  in order  to avoid arriving to  contradictions. The usage was later extended beyond that original notion.}  implying that one should only work within
one sector of the theory (e.g. a putative value for $\alpha$ in
the example above).

\vskip.5cm{\em{\bf Observation 2:} There can be regions in
parameter space where the ground state becomes degenerate, yet
symmetric ground states can always be found among the set of
degenerate ground states. } \vskip.5cm

A similar phenomenon will take place in field theoretical systems.
In fact we will show how the ground state of QFT's with CSSB is
indeed symmetric. However, the energy of certain non symmetric
states approaches that of the symmetric ground state in the limit
where the spatial extension of the system becomes infinite. In
that limit, the vacuum becomes degenerate and non symmetric ground
states exist; however, the symmetric ground state is always there
for all values of the spacial extension of the system.


 Next we consider the analogous issues in the field theoretical context.

\section{Continuous Symmetries  in Field Theories }
\label{Cont}

The simplest example of spontaneous symmetry breaking in field
theories is perhaps the linear sigma model.  This is a theory of
an n-tuplet of scalar fields $\Phi_i$, $ i=1,...n$, with a global
$O(n)$ symmetry, with Lagrangian density; \be\label{lsm} {\cal L}
= (1/2) \sum_{i=1}^n \partial_{\mu}\Phi_i \partial^{\mu}\Phi_i +
V( \sum_{j=1}^n\Phi_j\Phi_j),
 \ee
 The symmetry of the theory is $
\Phi_i \to R_{ij}\Phi_j$ where $R_{ij}$ is an orthogonal $n\times
n $ matrix. When the potential has the typical Mexican hat form
$V= \frac{\lambda }{4 }(\Phi_i \Phi_i -v^2)^2 $,  one might expect to
have a spontaneously broken symmetry situation, i.e.,  to have a
ground state which does not share the symmetries of the theory.
Indeed it is usual to see the ground state $|0\rangle$
characterized by \be\label{noway} \langle 0|\Phi_i|0\rangle=0,
\qquad i=1,.. n-1, \qquad\langle 0|\Phi_n|0\rangle=v
 \ee
 which is clearly not
invariant under the $O(n)$ rotations. Is such symmetry breaking
state the true  ground state of the system? In what follows we show that
this is not the case.

For the sake of simplicity, and definiteness, let us consider the
case of a theory of two scalar fields so $n=2$, and the standard
Mexican hat potential $ V= \frac{\lambda }{4 }(\Phi_1^2 + \Phi_2^2 -v^2
)^2$ (the general case can be dealt with in a similar fashion). In
this case it is convenient to introduce a new parametrization of
the fields by writing $\phi_ 1 = (\rho +v) \cos ( \theta)$, $\phi_
2 = (\rho +v) \sin ( \theta)$, leading to a 
Lagrangian:
 \be
  {\cal
L} = \frac{1}{2} \partial_{\mu}\rho
\partial^{\mu}\rho+ \frac{1}{2} (\rho +v)^2  \partial_{\mu}\theta
\partial^{\mu}\theta -\frac{\lambda}{4} (\rho^2 + 2 v \rho )^2 
\ee
The conjugate momenta are $ \pi_\rho = \dot \rho $ and $\pi_\theta
= (\rho +v) ^2 \dot \theta $. The Hamiltonian  density is then:
\be 
{\cal H} = \frac{1}{2} \pi_\rho^2 + \frac{1}{2 (\rho +v)^2}
\pi_\theta ^2 + \frac{1}{2} (\partial_{i}\rho)^2 + \frac{1}{2}
(\rho +v)^2 (\partial_{i} \theta)^2 +\frac{\lambda}{4} (\rho^2 +
2v \rho )^2 
\ee 
This can be separated into a free part and an
interacting part, with the free part given by \be {\cal H}_{f} =
\frac{1}{2} \pi_\rho^2 + \frac{1}{2 v^2} \pi_\theta ^2 +
\frac{1}{2} (\partial_{i}\rho)^2 + \frac{1}{2}  v ^2 (\partial_{i}
\theta)^2 +\frac{1}{2} m^2 \rho ^2.  \ee where $m^2 = 2 \lambda
v^2$.
 The free theory can now
be quantized in the standard fashion taking care of the special
treatment for the zero mode corresponding to the space independent
value of $\theta$. To make things more transparent let us consider
space to be a finite box of side $L$ and impose periodic boundary
conditions\footnote{We will  investigate also
  what happens in the limit $L\to \infty$ but note that
the finite size of the box has no bearing on the infinite number
of d.o.f. which is usually called upon as part of the explanation
for the differences between field theory and
  the  simple case of standard  quantum mechanics.}.
Before proceeding it is convenient to make a canonical
transformation to the new variables
\[\phi = v \theta\ \ \ {\rm and}\ \ \
\pi_ \phi = (1/v) \pi_\theta .\]
 The  free Hamiltonian  is therefore:
 \be
  H_f = \int d^3x
\frac{1}{2} \pi_\rho^2 + \frac{1}{2 } \pi_\phi ^2 + \frac{1}{2}
(\partial_{i}\rho)^2 + \frac{1}{2}  (\partial_{i} \phi)^2 +\frac{1}{2} m^2
\rho ^2 ]  \ee
 Here we must recall that this description is appropriate under the
 assumption that the interactions can be treated as a
 perturbation  and as long as the fields do not leave the
 corresponding  ranges implicit in their definitions i.e. $ \rho \in (-v, \infty)$,  and
with the understanding that there is an implicit identification of $\phi$  and
$\phi + 2n \pi v$.

\subsection{The vacuum state in a (spatially) compact universe}\label{conta}

As mentioned in the introduction, the spontaneous symmetry
breaking phenomenon is presented as a feature of field systems of
infinite extension. In order to study such statement in detail it
will  be convenient to define the field theory for a finite universe
and study the limit in which the infrared cut-off is removed.
Consequently, in this subsection we will analyze  with care the free part of the  linear sigma
model introduced above,  focussing on the situation  where the field theory is defined
on a {\it  finite}  three dimensional square box of size $L$ with periodic
boundary conditions.

We expand the fields in Fourier components writing
 \be \nonumber\rho(\vec x, t) = \sum_{\vec k}
\frac{(a_k (t) e^{ -i\vec k \vec x} + a_k^\dagger (t) e^{ i\vec k \vec
x})}{\sqrt{2 \omega_k L^3}}, \ \ \pi_\rho (\vec x, t) =-i \sum_{\vec k}
\sqrt{\frac{\omega_k}{2L^3}} (a_k (t) e^{- i\vec k \vec x} - a_k^\dagger (t)
e^{ i\vec k \vec x}) 
\ee 
where $\omega_k = \sqrt{ \vec k\cdot \vec k + m^2} $
and $[a_k,a^{\dagger}_{k^{\prime}}]=\delta_{k,k^{\prime}}$.  Here we have
ignored the bounds of the range of $\rho$, because the mass associated with that
field indicates that large departures from its zero value would require very
large energies\footnote{Note that this holds
for small fluctuations of either sign.}, and thus could be safely ignored when interested in features
of the ground state and nearby states. The sum is over the $\vec k$ with
components $k_i$ of the form $ k_i= 2n_i\pi/L $ for $n_i\in \Z$. Similarly we
write the Fourier expansion for the field $\phi$ except, that in this case
there are two crucial differences: First and foremost, special care must be
taken of the zero mode which needs to be treated separately, but second we
must recall that the range of the field variables is limited to $(0, 2\pi v)$,
and given that this field is massless, one can not rely on arguments as simple
as those employed when considering $\rho$. This is a fact that will have to be
kept in mind when attempting a mode decomposition.  We start by writing a mode
decomposition of the field and momentum conjugate: 
\be \phi =
\frac{\phi_0(t)}{ L^{3/2}} +\frac{1}{ L^{3/2}}\sum_{\vec k\not=0} \phi_k (t)
e^{ i\vec k \vec x}, \qquad  \pi_\phi =\frac{1}{ L^{3/2}} \pi_0(t) +\frac{1}{
L^{3/2}}\sum_{\vec k\not=0} ( \pi_ \phi )_k (t) e^{ -i\vec k \vec
x}\label{dieci} \ee
The Hamiltonian is then: \be H_f=\int {\cal H}_{f}\ d^3x =
\frac{1}{2 } \pi_0^2 +\sum_{\vec k}[ \frac{1}{2}(( \pi_\phi)_k
)^2+\frac{k^2}{2} (\phi_k )^2 ] +\sum_{\vec k}[ \frac{1}{2}((
\pi_\rho)_k )^2+\frac{k^2 +m^2}{2} ( \rho_k )^2 ]
 \ee
 We see that
this is the Hamiltonian of one free particle together with
an infinite collection  of harmonic oscillators. This view is
deceptive, because of the constraints associated with the  finite range of
the variable $\phi (x)$ which, in terms of the mode decomposition
leads  among other things  to  $\phi_k \in [0, 2\pi v L^{3/2}] $. We call this
 issue  (especially the nontrivial constraints  involving collectively the multiple modes) the ``range constraint"  (RC).
 It should be clear that the  precise treatment  of  the constraints  would be  a rather  complicated
 analysis involving all the modes, except the zero 
 mode\footnote{
  The exact nature of
these constraints is that if two field configurations $\phi_1(x)$ and $\phi_2(x)$
differ by $2\pi n(x) v L^{3/2}$, where $n(x)$ is an integer-valued function, then the
wave functional should satisfy  $\Psi [\phi_1(x)] =\Psi [\phi_2
(x)]$. This condition implies complicated constraints for the modes in
general. However, taking $n(x)=$constant one gets that the zero mode should be
in the range  $[0, 2\pi v L^{3/2}]$. The constraints on the other modes are independent of the
value of the zero mode if its amplitude is constant (as in the case of the
vacuum state considered below).}. 
This is an important
point, the zero mode  has a bounded range but is otherwise
completely decoupled, in the constraints,  from the other modes.

Let us assume for the moment that we can ignore the
(non-perturbative)
complications associated with the RC (for instance by considering only small amplitude excitations  for all  modes). This assumption will
be justified below.
{\bf The first thing we  must note is that   we find the mass-less scalar excitations or Goldstone bosons  usually associated with SSB even though the vacuum  is, as we will next show,  symmetric}.  Furthermore, we see that in
   that case we could write for all the modes, except  for $ \phi_0$, the
   standard harmonic oscillator creation and anihilation operators,
   namely
    \be
     \nonumber \phi = \frac{1}{ L^{3/2}}\phi_0(t) +\sum_{\vec k\not=0}
   \frac{(b_k (t) e^{ -i\vec k \vec x} + b_k^\dagger (t) e^{ i\vec k
   \vec x})}{\sqrt{2\omega'_k L^3}} \ \ \pi_\phi =\frac{
   \pi_0(t)}{ L^{3/2}} -i\sum_{\vec k\not=0}
   \sqrt{\frac{\omega'_k}{2L^3}} (b_k (t) e^{- i\vec k \vec x} -
   b_k^\dagger (t) e^{ i\vec k \vec x})\label{ff}
   \ee
   where
   $\omega^{\prime}_k = |\vec k| $,
and the  Hamiltonian becomes
 \be H_f=\int {\cal H}_{f}\ d^3x =
\frac{1}{2 } \pi_0^2 +\sum_{\vec k} \omega_k( a_k^\dagger a_k
+\frac{1}{2} ) + \omega_k' ( b_k^\dagger b_k +\frac{1}{2} ).
 \ee
  It is easy to see that the state  with minimal energy will be characterized by $a_k|0\rangle=0,
 b_k|0\rangle=0$,  and $\pi_0|0\rangle=0 $.
 We can  now write the
 corresponding  vacuum state wave functional, i.e., the minimum
energy eigenstate of the Hamiltonian, as \be \label{vacc} \Psi_0
=\prod_{k} \psi^{\phi}_{k}[\phi_k] \psi^{\rho}_{k}[\rho_k], \ee
where \be \psi^{\rho}_k[\rho_k] =N_k\ {\exp[-\frac{\sqrt{\vec
k^{2}+m^2}\rho^2_{k}}{ 2}] }
, \ee
 and
 \be
\psi^{\phi}_k[\phi_k]=N'_k \ {\exp[- \frac{\sqrt{\vec k^{2}}
\phi^2_k }{2}]}
, \ \ \ \mbox{for all $\vec k\not=0$ while $\psi_0^{
\phi}[\phi_0]=\frac{1}{(2\pi v L^{3/2})^{1/2}}$,} \ee
where $N_k$
and $N'_k$ are normalization factors.  The last equation shows
that the zero mode of the field has a constant wave functional.
Despite of the composite nature of the operators $\phi_1=(\rho +v)
\cos ( \phi/v)$, and $\phi_2=(\rho +v) \sin ( \phi/v)$, one can
easily prove that due to the contribution of the zero mode, their
expectation values in the above vacuum state, vanishes for any  finite value of the (in principle necessary) UV cut-off $(L)$, namely, that,
\be
\label{key1} \langle0|\phi_i|0\rangle=0 \ \ \ i=1,2.
\ee
 This
equation shows that the symmetry is not broken by the vacuum in
contrast with the naive expectation (\ref{noway}). Moreover, one
can explicitly show that
\be\label{key2}
\exp(i\epsilon\int
\pi_{\phi}(x) dx )|0\rangle=|0\rangle, \ee where the operator acting on
the left is the $U(1)$ rotation operator with
angle $\epsilon$ around the bottom of the Mexican hat potential.

The symmetric ground state (\ref{vacc}) has zero charge $\pi_0
\Psi_0=0$. Note that the symmetry implies the conservation of
$\pi_0$, namely $[\pi_0,H]=0$ where $H$ is the full Hamiltonian.
Therefore, no process in a universe, preserving the symmetry of the
Lagrangian (\ref{lsm}), would be able to change or be sensitive to
the energy stored in the zero mode\footnote{ We should note however
that   although,  we  do not expect it to  affect the
symmetry, gravitation is  expected to be  sensitive to the
energy of the quantum  field.}. The eigenvalues $\pi_0$ would label, in this
sense, disjoint super-selection sectors (see footnote \ref{ssrules}). Nevertheless, even if we
take another eigenstates of $\pi_0$ the symmetry would not be broken
in any sense as the value of the field $\phi$ is completely
undetermined due to the uncertainty principle---alternatively, the
probability density $\rho(\phi_0)$ is constant and all $\phi_0$ are
equally likely. Conversely, there is no process in a symmetry
preserving universe that can prepare a state in a symmetry violating
configuration starting from  a symmetric state.

Note that in order to satisfy equation (\ref{noway}) one would have
to modify the zero mode wave function from a constant to some
localized packet. For instance, one could take $\psi_0^{\phi}=(\pi
\sigma^2)^{-1/4}\exp[- \phi_0^2/\sigma^2]$, for some chosen value of
the dispersion $\sigma$. This would clearly increase the expectation
value of the energy at least by the amount $\Delta
E=\langle\pi_0^2/2\rangle \approx \sigma^{-2}/2$. A state satisfying
(\ref{noway}) is an excited state. Essentially the reason for the
energy difference between the two states lies in the extra
contribution from the momentum terms to the localization of the wave
packet along the bottom of the potential.  It is thus clear that all
things being equal the state without that energy contribution will
have lower energy.  That is, the true vacuum of the
theory\footnote{This point is not affected by the fact we can not
describe the state  fully due to the nontrivial constraints (RC) present among
the modes $\phi_k$.} is one for which the zero mode is described by $\pi_0=0$, i.e.
a wave functional which is uniformly spread over the bottom of the
potential well, and thus fully symmetric, i.e. the state
(\ref{vacc}) above.

Leaving aside for the moment the question of the preparation of
symmetry violating state, one could in fact consider a
construction of perturbation theory based on a state which is `non
symmetric', i.e., where the zero mode is initially sharply peaked
about a certain angle, and thus is not the true vacuum. That would seem to correspond to what is
done in the usual treatments of this problem; however, evolution
will destroy the initial sharp asymmetry of the zero mode wave
function.  This can be explicitly checked using elementary quantum
mechanics of a free particle on a circle of radius $v$. Indeed if
we start with a state for $\phi_0$ localized somewhere around the
circle with spread $\sigma_{\phi_0}$,  then, after a time of the
order of $t_0=\hbar^{-1}(2\pi v L^{3/2}
\sigma_{\phi_0}-\sigma^2_{\phi_0})$ the original localization is
completely lost. Of course, such
 a state   does break the symmetry. However, it does so in  weaker sense as
no special value of $\phi_0$ is sharply selected  at all times by
the corresponding  probability distribution. Therefore, if we
consider the zero mode as a quantum degree of freedom there is no
way to make sense of a  stationary  state satisfying
(\ref{noway}).

Moreover, if we include the gravitational interaction in our
considerations,  recall that without breaking  the symmetry of
(\ref{lsm}), gravity is sensitive to the  total  value  of energy
of the system,  rather than to energy differences. In other words,
unless the zero mode is in the symmetric state (\ref{vacc}),
there would  be excess energy stored in the field which would
behave as a non trivial source of gravity, and  which is
therefore,  in principle, detectable.

Finally,   when considering a non-symmetric state, we must keep in mind that any interaction, no matter how
small, that violates the symmetry (and  which is not taken into
account in (\ref{lsm})) would tend to drive  the system in the
direction of  relaxing to the true vacuum (\ref{vacc}).  The
extent of this relaxation would of course depend on the full
details of the initial state of the entire multi-system, and of
the interactions between its parts.

Incidentally,  we should note that the  issues associated with the
ranges and constraints in $\phi_k$ cannot be ignored.  In fact
the vacuum state above, has for the variable $\phi_k$ a width $\sigma_k =
{1/\sqrt{\va |\vec k|}}$, and that would have the support of the
wave function exceeding the range of variable $\phi_k$ for $|\vec
k | <[(2\pi)^2 v^2 L^{3}]^{-1}$. This would not seem to be a
serious problem for nonzero modes, in any relevant situation.
Recall that $\vec k =\frac{2\pi}{ L} (n_1, n_2, n_3)$ so that the
problematic condition would correspond to $n_1^2 + n_2^2 +n_3^2 <
[(2\pi)^3 L^2 v^2]^{-2}$ which for large enough $L$ would not
apply to any of the non-zero modes. Therefore, the perturbative
assumption made here (i.e. to ignore the RC) is consistent as long
as we take a sufficiently large box.

\subsubsection{The infinite size limit}

Let us now consider the infinite size limit i.e. $L\to \infty$.
The first observation is that the localization of the wave
function in the variable $\phi$ is not the relevant thing to
consider as the value of the field is characterized by the
variable $\theta$  whose range is $[ 0, 2\pi] $.  Localizing the
corresponding  global  variable in the narrow band $ \delta\theta$
corresponds to localizing the variable $\phi$ in the range $
\sigma =\delta\theta v L^{3/2} $  and that would increase the
energy of the  state by an amount $ E = \frac{1}{ 2v^2
L^3(\delta\theta)^2}$. This quantity goes to zero as $L \to
\infty$,  as long as $(\delta\theta)$ does not go to $0$ as
$1/L^{3/2}$ or faster. This indicates that  in the limit, we are
lead to a situation in which the ground state is highly
degenerate.  That is,  as candidates for lowest energy state, we
have, on the one hand, the symmetric state, and on the other, a
large collection  of  asymmetric states where the zero mode is
localized to an arbitrarily finite degree  on any arbitrary point
in the range $\theta \in [0, 2\pi]$, and in fact many others.

The second observation is that the vacuum associated with the Fock
construction about any of these states  would  correspond to a
wave packet  concentrated,  within the allowed range  for the
variable $\phi_k$,  only  for those values of  $k$   such that
$|\vec k |>[(2\pi)^2 v^2 L^{3}]^{-1}$.  However,   in the limit $
L\to \infty$ they would  include  all the values of   $\vec k
\not= 0$.

However nothing of what we have said so far indicates which of the
vaccua should one use as the ``true vacua". The symmetric state is at
least as good a state, as the others, and it is in fact, the one that
corresponds to the true vacua for the cases where the value of $L$ is
finite. The degeneracy appears only as the result of the infinite limit
on the size of the box.  However, in the infinite size limit, among
the degenerate set of lowest energy states one might choose a state
which breaks the symmetry. The situation is just the analog of the one
considered in the QM example that motivated the Observation 2 in
Section \ref{QM}.

 Finally say one wants to focus  on  the issue of which  one of
those  states  is most appropriate to use when attempting to
represent the  ground state of a particular  system: in so doing
we need to face several issues. First and foremost  at the
practical level, it is clear that when dealing  with a {\it
localized experiment}, the  size of our universe should not matter
and  one might be tempted to conclude that  for all practical
purposes it could be taken to be infinite. However we must be
careful about how this limit is to be taken:
\begin{enumerate}
\item   should  we consider the expression for the  Hamiltonian with finite $L$, then take  the limit,
and then find its eigenstates and in particular its vacuum?,  or

\item  should we consider the  expression for the  Hamiltonian with finite $L$,  then find its eigenvalues and eigenvectors,
and in particular the vacuum,
and then  consider the limit when $L$ goes to infinity?
\end{enumerate}

This is a  very relevant question as it is  clear, from both, the
example above and that discussed in the context  of Observation 2,  that  the  two procedures  do not always  commute.
The answer lies in the fact that what we have, in general,  in
such situation is a system with  finite extent and for which we
want to use  the simplifications that  arise by considering one of
the length parameters of the problem to be   much larger than all
the others, and thus whenever the two alternatives are different
we must recognize that the realistic situation corresponds  to the
second one.
From the above  discussion,  it  seems  clear that the possibility
of a vacuum  that  does not share the symmetries of the theory,
is, in this,  the most common field theoretical context  in which
SSB is considered,  associated not only with a system with
infinite degrees of freedom, but a system with infinite spatial
extent. Moreover we have seen that the analysis  of the
idealizations that lie behind the standard approach indicates that
the consideration of  SSB in  the case of infinitely extended
systems (i.e.  the selection of  an asymmetric   vacuum state
within option 1) is not justified when its conclusions differ from
those  one arrives to  in taking option 2).

\subsubsection{The zero mode, causality and clustering}

We have seen that the question of spontaneous symmetry breaking is
intimately related with the behavior of the wave functional of the
zero mode. However, as shown by the above argument some of the
physical contributions of the zero mode (e.g. the contribution to the
energy density) become negligible in the large volume limit. Does that
mean that we can simply neglect that mode in the quantization and
consider it fully classical? If so, then the question of symmetry
breaking will become a classical question concerning the zero mode
position around the bottom of the mexican hat potential.  In such
context condition (\ref{noway}) would be a clear-cut semiclassical
statement. In this subsection we show that (in the spatially compact
case) the zero mode cannot be neglected in the quantization
process. Doing so would be in direct conflict with local causality.

In order to see this explicitly let us compute the commutator
 \be
 \label{commutatorE}
[\phi(x),\phi(y)]=\sum_{\vec k\not=0} \frac{1}{2 \omega'_k
L^3}\exp (-ik\cdot(x-y))-\sum_{\vec k\not=0} \frac{1}{2 \omega'_k
L^3}\exp (ik\cdot(x-y))+2i \frac{(y^0-x^0)}{L^3},
 \ee
 where the
last term comes from the contribution of the zero mode, namely
\[
[\phi_0(x^0),\phi_0(y^0)]=[\phi_0(x^0),\exp(i(y^0-x^0)\pi_0^2)\phi_0(x^0)\exp(-i(y^0-x^0)\pi_0^2)]=2i(y^0-x^0).
\]
In the non compact case the sums become integrals, and  the functional form
of the commutator is in that limit
$[\phi(x),\phi(y)]=D(x-y)-D(y-x)$. Invoking  Lorentz invariance
of the function $D(x)$, one can easily see that the commutator
vanishes if $x-y$ is space-like. In the compact case the
commutator is still zero when $x-y$ is space-like (as expected from
causality). However, the Lorentz invariance of the sums in the
expressions above  can not be warranted  because Lorentz invariance, in this situation,   is broken by the boundaries, 
 and in fact  an explicit calculation  indicates that $D(x)$ is no  longer  invariant and that the causality
requirement is satisfied only  thanks to the precise cancellation of the problematic term resulting  from the
sums and the contribution to the commutator coming from the zero
mode\footnote{The simplest way to explicitly verify this is to
work in $1+1$ dimensions. Taking $L=1$ space-time events are
labeled by $x=(\varphi,t)\in S^1\times \R$ and $k=n\in \Z$.
 We
have the following identities for the first series in
 (\ref{commutatorE}):
\ba\nonumber && \frac{1}{2 \pi}\sum_{k\not=0} \frac{1}{2 k} \exp
(-ik\cdot x)-\frac{1}{2 \pi}\sum_{k\not=0} \frac{1}{2 k} \exp (ik\cdot x)=\\
\nonumber && -\frac{1}{2 \pi} \sum_{n=1}^{\infty} i \frac{\sin(n
(\varphi+t))}{n}- \frac{1}{2 \pi}\sum_{n=1}^{\infty} i
\frac{\sin(n (-\varphi+t))}{n}=-\frac{1}{4\pi}\log{\frac{1-e^{-i
(\varphi+t)}}{1-e^{i (\varphi+t)}}} +
\frac{1}{4\pi}\log{\frac{1-e^{-i (-\varphi+t)}}{1-e^{i
(-\varphi+t)}}}=\\ \nonumber &&=\frac{it}{2
\pi}+\frac{1}{2}\left({\rm Int}(\frac{t+2\pi}{2\pi})+{\rm
Int}(\frac{t-2x+4\pi}{2\pi})\right),\ea where ${\rm Int}(\cdot)$
denotes the integer part function. With all this we see that the
zero mode contribution cancels the term linear in $t=x^0-y^0$ and
$[\phi(x),\phi(y)]$ vanishes at space-like separated points.}. The
point is that the zero mode (which is the  one that carries the
information about the symmetry of the vacuum) is a genuine quantum
degree of freedom for any value of the volume and must be
treated accordingly  in  considering the infinite volume limit.

However, if the zero mode of the chosen vacuum state is in a
quantum state with non vanishing spread in $\phi_0$ the clustering
property is violated. This can be seen from the computation of the
equal time correlation function $\l \phi(\vec x,t)\phi(\vec
y,t)\r$, namely, using equation (\ref{ff}) we have
\[
\l \phi(\vec x,t)\phi(\vec y,t)\r=\l \frac{1}{  L^{3}} \phi^2_0
+\frac{1}{ L^{3}}\sum_{\vec k\not=0} b_k b^{\dagger}_{k}
\exp{(-i\vec k\cdot \vec x)} \exp{(i\vec k\cdot\vec y)}\r=\frac{1}{
L^{3}} \l\phi^2_0\r + \delta(\vec x,\vec y)
\]
 which means that the
correlation
 \be \Delta(\vec x,\vec y)=\l \phi(\vec x,t)\phi(\vec
y,t)\r -\l \phi(\vec x,t)\r\l\phi(\vec
x,t)\r=\l\phi^2_0\r /L^3+\delta(\vec x,\vec y)-{\l\phi_0\r}^2 /L^3
=\sigma^2_{\phi_0}/L^3+\delta(\vec x,\vec y)
\ee
 which does not
depend of $x$ and $y$ as long as the points are different. Here
$\sigma_{\phi_0}$ denotes the width of the wave function of the
zero mode.
 We note that if the zero mode is in the minimum energy
state, i.e., the wave function $\Psi(\phi_0)=1/(2\pi v L^{3/2})^{1/2}$
(recall that from its definition (\ref{dieci}) one has $0\le
\phi_0\le 2\pi v L^{3/2}$) then $\sigma^2_{\phi_0}=(1/3) \pi^2 v^2
L^3$.
Hence, in the true vacuum there are long distance
correlations given by
 \be
 \Delta(\vec x,\vec
y)=\frac{\pi^2v^2}{3}+\delta(\vec x,\vec y). \label{indy}
\ee
 Thus the clustering property is violated independently of the size of
the universe.  Notice that if we were to consider the lack of
clustering as a problem we must be aware that it will not be solved by
considering the translationally invariant state with sharp orientation
in $\phi_0$ (the quantum counterpart of the state usually taken as
the emblematic state of SSB) as the ground state, because in that case
the vacuum will violate the clustering property when considering the
n-point functions of $\pi(x)$ rather than $\psi(x)$: In fact it is  easy
to see that for a state which has a width $\sigma^2_{\pi_0}$ in the
variable $ \pi_0$, one has: \be\nonumber \Delta_{\pi}(\vec x,\vec
y)=\l \pi(\vec x,t)\pi(\vec y,t)\r -\l \pi(\vec x,t)\r\l\pi(\vec
x,t)\r=\l\pi^2_0\r /L^3+\delta(\vec x,\vec y)-{\l\pi_0\r}^2 /L^3
=\sigma^2_{\pi_0}/L^3+\delta(\vec x,\vec y) \ee and, the Heisenberg
uncertainty relation between $\psi_0$ and $ \pi_0$ (note that these
two quantities have standard commutation relations) warrantees that a
sharp value of the former requires a large value of the uncertainty of
the latter.  Therefore,  we must conclude that the clustering violation exhibited here--- and in particular
the $L$-independent long distance correlation (\ref{indy}) produced by
the symmetric ground state (\ref{vacc})---is a clear-cut manifestation of the
quantum nature of the zero mode that survives the infinite size limit.

\subsection{The vacuum state in a (spatially) non-compact universe}

We have illustrated with a simple model how no spontaneous symmetry
breaking is possible in QFT's defined on spatially compact
space-times. The only way to have the ground state breaking the symmetry
of the system would be to have a zero mode that behaves classically.
In the compact case the latter possibility is inconsistent as it
would lead to conflict with local causality. However, the situation
changes drastically in the non-compact case where the  technical necessity  to
 impose  certain boundary conditions  in the  mathematical treatment of 
 problem, `freeze' the zero mode at some position around the Mexican hat potential.  That is,  in this treatment the
boundary conditions requiring (quantum) fields to decay at spatial
infinity, simply drop the zero mode from the set of (quantum) degrees
of freedom from the onset. Consequently, the ground state of such model breaks the
symmetry in a way that is mathematically in strict correspondence with the CSSB
scenario: what breaks the symmetry is the localization of the zero
mode which is a classical degree of freedom.  The view point taken in
such treatment is essentially perturbative in nature. In other words,
one quantizes fluctuations around some classical solution of the field
equations which violates the symmetry from the beginning: the homogeneous
field configuration defined by a point on the bottom of the Mexican
hat potential in our example.

In view of our discussion of the previous sections it is clear that
the symmetry breaking being considered here is introduced by our
mathematical limitations concerning the treatment of quantum fields in
a non-compact space. Indeed, one could decide to model the physics
that is tested in the lab by assuming that the universe is spatially
compact (yet very large). It is clear that such model will introduce
all sorts of complications of practical nature; it will for instance
make the definition of scattering theory rather involved. However, it
should be clear that these formal  difficulties are irrelevant from
the physical view point, and that for sufficiently large spacial
dimensions all the local phenomenology of, say,  particle physics, could be
recovered. Therefore, ignoring some  practical difficulties, the
alternative treatment in terms of a compact space,  indicates that the symmetry is retained by the
ground state. Thus, our view  is that in the standard  treatments  one is  not  really  justified  in stating
that the symmetry
is broken, because the symmetric states have been removed  from  consideration  by our
boundary conditions (the latter being a choice of the physicist
and not of nature).

In the references \cite{Nair} and \cite{Jackiw:1995be} the authors formally 
argue that the state \[|\epsilon\rangle\equiv \exp(i\epsilon\int
\pi_{\phi}(x) dx ) |0 \rangle\] is in fact orthogonal to $|0\rangle$: a manifest
illustration of the breaking of the symmetry by the ground state.
This is done by using a ``regulated'' expression for the integral 
in the argument of the above exponential, namely:
\[\pi^{\va (L)}_0\equiv \int
e^{-\frac{x^2}{L^2}}\pi_{\phi}(x) dx. \]
We would like to note that the formal character of such derivation 
is very misleading.

We should warn the  reader of  the fact that the situation that
is envisaged in the discussions  seeking  justification of the
spontaneous breaking of the symmetry, has some built-in  aspect
that  seems  rather suspect,  as, on the one hand it  requires to
treat the system as having infinite spatial extent (not just
infinite number of d.o.f. as shown in the previous subsection),
but at the same time, demands us to restrict ourselves to the
algebra of local operators, by arguing that in practice all of our
laboratories and apparatuses are of finite extent. One is,  {\it
ab initio}, setting up  the problem in a way where the zero (or
homogeneous) modes of the system, would end up being  treated
classically.  We have seen that this is not justified by physical
considerations, and not needed for the construction of the quantum
theory.

 Finally we note, that in fact, by showing that  Nambu-Goldstone bosons do emerge
under the conditions usually thought to be associated with SSB even if the
vacuum is symmetric, the analysis we have presented opens a path for
reconsidering the implications of the so called Mermin-Wagner-Coleman theorem
\cite{Coleman}. What the theorem actually shows that, in 1+1 dimensions, the
vacuum state must be invariant under any continuous symmetry of the
action. This is often interpreted as implying the absence of Nambu-Golstone
bosons.  As we have shown, it is generically the case that the vacuum is
invariant under the continuous symmetries, and that this is by no means in
conflict with the existence of Nambu-Goldstone bosons associated with the
symmetry. What is special about two dimensions is that there is no way of
making sense of a mass-less scalar field in two dimensional Minkowsky
space-time\cite{NO 2 D SFT}.  {The problem is closely related to infrared
divergences that seem impossible to regularize in any sensible way. These
divergences imply vacuum must be invariant in \cite{Coleman}. The present work
indicates that the option of Nambu Goldstone bosons (in a theory that is not
free) when the vacuum is invariant, has not been ruled out by that theorem. On
the other hand, there is closely related situation (which could be consider as
an IR regularization of the former although with a breaking of Lorentz
invariance) that clearly allows Nambu-Golstone exitations: the spacially compat model studied in
Section \ref{conta}.  In such case, the vacuum is perfectly symmetric and 
mass-less Nambu-Goldstone bosons
are present.}


 \subsection{ The case of approximate symmetries.}\label{aps}
 
  It is clear that if the symmetry one is considering is only an approximate
  symmetry, a big part of the analysis we have been making throughout  this
  manuscript would not longer apply.  On the other hand, physicists in various
  fields do often consider symmetries that are only approximate symmetries of
  the system they want to study. The reason for that is the idea that one can
  consider, as a starting point of the analysis, the case where the symmetry
  is exact, which is in fact often easier to study than the realistic
  situation, and where the latter is regarded as involving the ``small
  corrections" associated with the say, Lagrangian terms, that break the
  symmetry and are to be treated by a suitable perturbative analysis.  Perhaps
  the most famous of these cases corresponds to the so called ``chiral
  symmetry" in the light quark sector of the standard model.
  
 In considering these situations we should be careful to avoid confusing the
 sequence in which the analysis is carried out with any sort of implicit time
 sequence describing when things do  occur in nature.
 
  In order to discuss these situations it is convenient to have a specific
 example in mind, and to that end we consider the model of Section II were we
 add to the Lagrangian a small perturbation that explicitly breaks the $O(n)$
 symmetry of that example.  We take for instance a linear term in one of the
 fields, such as $ a^3 \Phi_1$ where $a$ has dimension of mass and a value
 that is small as compared with all other nonzero mass scales of the system.
 It is clear that this modification leads to potential that resembles a
 mexican hat but with a slight tilt.  What used to be the flat bottom of the
 potential, has now acquired a small inclination. The standard way to approach
 this problem is to start off by disregarding the small term that explicitly breaks the symmetry, 
 and to study the symmetric case which has the characteristics
 that are thought to lead to the phenomena of SSB.  Among these we have, for
 instance, the emergence of mass-less Nambu-Goldstone bosons. The perturbation
 analysis then indicates that the symmetry breaking term would give
 the Nambu-Goldstone bosons a small mass characterized by the parameter $a$.
 As a result of the full analysis one ends up with a picture where
 the vacuum is not symmetric and where  there is a set of very light particles called
 pseudo-Nambu-Goldstone Bosons, which, in the case corresponding to the
 chiral symmetry mentioned above, are identified with the light mesons
 (specifically the three pions, corresponding to the $SU(2)$ chiral symmetry of
 the pair of almost mass-less quarks $u$ and $d$, or   to the $SU(3)$ chiral symmetry leading to the
 octet of light mesons involving also  the quark $s$), the analysis of which was
 recently awarded the physics Nobel Prize.

 The fact is, however, that as we have seen, one can carry out exactly the
 same type of perturbative analysis but starting out  instead with  the  recognition that after
 the identification of the approximate lagrangian (obtained by setting the
 parameter $a$ to zero), we have an exact symmetry and that the ground state
 is  thus symmetric\footnote{In particular we have shown that in the case where the
 box is finite ($L$ is finite), then there is no degeneration in the ground
 state, and that it is exactly symmetric.  It would be highly undesirable to
 take the position (of relying on the viable alternative of choosing by hand
 one of the asymmetric vacua, that we saw, do exist when the universe is
 infinite) that the size of the Universe has any bearing on whether the
 analysis of the  breaking of the chiral symmetry is correct or not. Indeed, relying on such
 posture one would be lead to conclude that as ``we know that the symmetry is
 spontaneously broken" the universe must be finite.}, as discussed in Section
 II.  This situation leads nevertheless to  the conclusion about the
 existence of  the corresponding Nambu-Goldstone bosons.  Once having treated this approximated
 description in a rigorous way, we can next use it as the starting point of
 the perturbation characterized by a nonzero value of  the parameter $a$.  This perturbative
 analysis would then lead to two very important conclusions: The first one is
 the acquisition of a small mass by the initially (in the sense of the
 analysis not of timing of events in nature) mass-less bosons, of a small mass
 characterized by the parameter $a$ and the second,  is a deformation of the
 vacuum, which will now be characterized among other things by the selection
 of the $\Psi_1$ field direction as the one characterizing the vacuum
 expectation value of the fields, in accordance with the equation \ref{noway}.
 The point is that we end again with the same final picture: non-symmetric
 vacuum, and light pseudo Nambu Goldstone Bosons, without ever coming into
 conflict with the overall analysis of this manuscript.  Moreover by doing so,
 we gain not only conceptually, but also in the understanding that the picture
 is independent of whether the universe is finite or not.
           
 At this point we would like to discuss  a simple  issue that might
 lead to confusion in distinguishing the physics associated to the limit $a\to
 0$ with that corresponding to the symmetric case $a=0$. For simplicity let us
 focus on the QM model of Section \ref{QM} where we add to the Lagrangean
 (\ref{unos}) a linear term $a X_1$ (this amounts to the analog explicit
 breaking of the $U(1)$ symmetry referred to above that slightly tilts the
 mexican hat potential along the $X_1$ direction).
The effective
    Hamiltonian for the angular degree of freedom in equation (\ref{cinco})
    acquires an extra term $ \delta  = a v \sin(\theta)$ which can
    be used to evaluate in first order perturbation theory the new ground
    state wave function for the variable $\theta$, which is given by $
    \Psi_{\theta} \approx \frac{1}{\sqrt{2\pi}} ( 1- 2 a m v^3 \sin(\theta)) +
    O(a^2)$, and which leads to a meaningful
expectation value $\langle \theta \rangle= 3\pi /2$.  Thus for large values of
    $\lambda$ and for $a$ small, we can write the ground state of the full
    asymmetric theory, approximately as $ \Psi'_0(r , \theta)
    =\Phi_0(r)\frac{1}{\sqrt{2\pi}} ( 1- a 2m v^3 Sin(\theta)) + O(a^2,
    1/\lambda)$, which can be contrasted with the ground state of the
    symmetric theory given by equation (\ref{wf0}). Thus we have (for $a>0$)
    that \be \langle 0|X_1|0\rangle=-v+O(a), \ \ \ \langle 0| X_2|0\rangle=0 \
    \ \ \forall\ \ \ 1>>a>0.  \ee Therefore, even when the ground state is
    symmetric at $a=0$, the limit $\lim_{a\rightarrow 0} \langle
    0|X_1|0\rangle=-v$, and $\lim_{a\rightarrow 0} \langle 0|X_2|0\rangle=0$
    yields non vanishing order parameters. Note that the wave functional of
    the ground state converges pointwise to the symmetric state (\ref{wf0})
    when $a\to 0$. The above situation is a prototypical example of non
    commutativity of integrals with limiting procedures 
\footnote{\label{Daniel}In order to separate the mathematical issue from the physics, consider the set real valued nonnegative
    functions on the line, viewed as probability distributions of the variable
    $x$ and the corresponding average $\l x\r$. 
For concreteness take a Gaussian distributions of width $1/a$ and centered at,
    say $ x=5$.  It is clear that the mean value $\l x \r=5$ independently of the
    value of $a$, so it follows that $\lim_{a\to 0}\l x\r=5$. However, the
    distribution corresponding to $a=0$ is the constant distribution for which
    the mean value of $x$ is not defined. In order to avoid this issue
    consider now distributions on a circle given by (periodic functions)
    $f(\theta)$ for $\theta \in [0,2\pi]$. Now the expectation value will be
    finite for any bounded $f(\theta)$. The analog of the previous example are
    for instance Guaussian distributions with center in $\theta =\pi$ and
    width $1/a$. As before $\langle \theta\rangle=\pi$ and $\lim_{a\to
    0}\langle\theta\rangle=\pi$. In the limit $a\to 0$, however the distribution is constant and
    thus gives to all values of $\theta$ the same likelihood, and it would make no sense
    to interpret this as indicating that somehow the symmetry under rotations
    in the circle is broken by the constant distribution.  The point of course
    is that in such case the uncertainty in the value of $\theta$ is equal to
    the full range of its possible values.}.
 It is clear
    that the above argument can be directly translated to the case of linear
    sigma model of Section \ref{Cont} and to any field theoretical situation
    where symmetries are only aproximate.  Therefore, the fact that our
    analysis implies the exact symmetry of the ground state in the symmetric
    world ($a=0$) is by no means in contradiction with the existence of non
    vanishing order parameters in a world with approximate symmetries (in the
    limit $a\to 0$).
 

An important example where the above discussion  applies is the famous case
 of  the chiral symmetry in QCD. Ignoring the details of the
electroweak Lagrangian, as well as the
heavy quarks, we can write an effective Lagrangian for the up-down sector as
follows:
 \be { \cal L }= -\frac{1}{4} {\rm Tr}[F_{\mu\nu} F^{\mu\nu}] +
\sum_{i=u,d} [ \bar\psi_R^{(i)} D_\mu \gamma^\mu \psi_R^{(i)} +
\bar\psi_L^{(i)} D_\mu \gamma^\mu \psi_L^{(i)}] + \sum_{i=u,d} m_i [
\bar\psi_R^{(i)} \psi_L^{(i)} + \bar\psi_L^{(i)} \psi_R^{(i)} ] \ee where
$F_{\mu\nu} $ is the Lie Algebra valued field strength tensor of the gauge
fields, $\psi_{R,L}^{(i)} $ are the right and left handed components of the
quark field with flavor $i$, $D_\mu$ are the color gauge covariant
derivatives, $ \gamma^\mu$ the Dirac matrices, and $ m_{i}$ the quark masses.
This Lagrangian has besides the Poincar\'e symmetries, the $SU(3)$ color gauge
symmetry (whose indices we are not explicitly shown for simplicity), and the
two $U(1)$ flavor symmetries.  If we set the two quark masses to the same
value the Lagrangian has an additional $SU(2)$ global flavor symmetry and if
we set these masses to zero it has an $SU(2)_L\times SU(2)_R$ where the first
term corresponds to the global flavor symmetry for the left handed components,
and the second for the right handed components.  In this later situation we
should have two $SU(2)$ Lie algebra valued conserved currents, which
is customary to write the combinations $J_V^{a \mu} (x )$ and $ J_A^{a \mu}
(x)$. The first stands for the current (called the vector current) is
associated with the simultaneous identical $SU(2)$ rotation of both the left
and right handed components, while the second (the axial vector current or
axial current in short) corresponds to the rotation of the left handed
components with the inverse of the rotation that is applied to the right
handed components.

The basic idea for the treatment of most problems in this area is to consider
the realistic case where $m_i \not= 0$ as a small perturbation from the
mass-less case which as we saw has an extensive set of symmetries.  This leads
one to write at the phenomenological level the well known expression
characterizing of the partial conservation of axial current (PCAC):
 \be
\langle 0|\partial_\mu J_A^{a \mu} (x) |k, b \rangle = - m_\pi^2 f_\pi
\delta^{ab} \psi_k(x) \ee where $ J_A^{a \mu} (x)$ is the axial current where
$ |k, b \rangle $ is a one pseudo-Goldston boson (pion) state with momentum
$k$, $b=1,2,3$ is an $SU(2)$ Lie algebra index, $m_\pi $ is the Pion mass,
$f_\pi $ is the pion decay constant, and $\psi_k(x)$ is the normalized wave
function for the scalar particle of momentum $k$.

It is otfen argued that when considering the behavior of the equation above in
the limit when the quark masses go to zero, there left hand side must vanish,
and that there are two options for the way the right hand side should
vanish. Either $m_\pi \to 0 $ or $ f_\pi \to 0 $.  The standard posture is to
say that is that $m_\pi \to 0 $ corresponding to the mass-less Goldstone
bosons characteristic of SSB, while $ f_\pi $ remains finite in this limit,
something that is associated with the breaking of the symmetry by the vacuum
of the theory. Indeed, one can characterize  the pion
decay constant as an order parameter of the form
\be f_\pi^2 \delta^{ab} = -\frac{i}{3} \int d^4x \langle 0| T (
J_L^a(x)\cdot J_R^b(0))
|0\rangle. \label{abo}\ee
The reason one calls the previous expression an order parameter is that it
vanishes if the vacuum is invariant under an axial transformation.
More explicitly, under an axial transformation $J_L^a$ is `rotated' as a vector in one
direction while $J_R^a$ is `rotated' the same amount but in the oposite
direction. Note that there is no non-trivial bi-linear invariant under
transformation that can be constructed in terms of the left and right
currents. Therefore, the operator whose expectation value 
we are computing in the previous expression has no singlet component under
axial transformations and its expectation value in symmetric states should 
vanish. 

Consequently, the fact that the
pion decay rate $f_{\pi}$ measured in the lab  is large (in comparison to the
light quark masses) could be (missleadingly) taken as an observational
evidence of the existence of a spontaneous symmetry breaking by the QCD
vacuum.  However, as we point out above even when the ground state (e.g. in a spatially compact universe) might
converge to a symmetric state in the limit $m_i\to  0$ the order parameters 
of the kind described in the equation (\ref{abo}) might remain large as the limit is taken.
    
{\bf  In other words, the limit
\be  \lim_{m_i\to  0} \left(- \frac{i}{3} \int d^4x \langle 0| T (
J_L^a(x)\cdot J_R^b(0))
|0\rangle\right) \label{abo2}\ee   which is  identified with the non zero quantity  $f_\pi^2 \delta^{ab}$  should not be confused with the evaluation of the LHS of  (\ref{abo})  in the theory  with $m_i=0$  which  does vanish.}
It is worthwhile pointing out that the uncertainties in the vacuum state of the
operator $\int d^4x T (
J_L^a(x)\cdot J_R^b(0))$ grow as the symmetry breaking parameters go to zero
(see footnote \ref{Daniel}).
As the light quark masses are very small, his may have some phenomenological implications whose study is outside the
scope of this paper. 

From the point of view discussed in this manuscript, the case of chiral
symmetry can be concisely described by saying that the explicit breaking of
the symmetry associated with the light quarks masses leaves an imprint in the
vacuum of the theory that does not disappear smoothly in the limit when these
masses are artificially made to tend to zero. The result is that the correct
results can be obtained either by starting of with an approximation where the
light quark masses are taken as zero and maintaining (without justification at
this level) the non-symmetric nature of the vacuum, and then using
perturbation theory to treat the nonzero value of those masses, or
alternatively, and more transparently, following the approach discussed
above.

We now turn  for completeness to  a  brief review  of the  situation of
spontaneous symmetry breaking in gauge theories, where, in
contrast to  what happened in the  case considered here,  we find
ourselves in complete agreement with the analysis found
in \cite{Symmetry Breaking}.

\section{ Gauge symmetries and the Higgs Mechanism}\label{gshf}

The situation of SSB of gauge theories could be described in a
single paragraph: as it is well known gauge degrees of freedom are
not physical. They are not accessible to any physical interaction
and are not governed by any equation of motion. They are simply
redundant fields that enter our mathematical description of
certain physical interactions. Consequently, a gauge symmetry
cannot be broken not only by the vacuum, but by any state in the
theory.  Nevertheless the  main  phenomenological features one
  associates  with a SSB in gauge theories (i.e. mass generation for gauge bosons
   and for  fermions)  remain valid despite this fact.
This statement resumes the conceptual content of this
section and the treatment of \cite{Strocchi}.

In those cases where the physical degrees of freedom can be
explicitly identified,  the illustration of the above statements
is obvious when expressing the dynamics in terms of gauge
invariant quantities. In this section we will study this issue in
the Abelian Higgs model where one encounters the Higgs phenomena
of mass generation for gauge fields.  The simplest example is
provided by a theory with a $U(1)$ gauge symmetry, gauge field
$A_{\mu}$ and a charged scalar field $\phi $ with Lagrangian
density;
     \be\label{hem}
   \sL_{AH} = D_{\mu}\phi {D^{\mu}\phi}^{*}+ V (\phi^{*} \phi)  + F_{\mu \nu}F^{\mu \nu}
   \ee
where $D_{\mu}\phi  = \partial_{\mu}\phi  + ig A_{\mu}\phi$ and
$F_{\mu \nu}=\partial_{\mu}A_{\nu} -\partial_{\nu}A_{\mu}$. The
gauge transformations of the  theory  correspond to  $ \phi \to
\phi e^{i g \alpha(x)}, A_{\mu} \to A_{\mu} -\partial_{\mu}
\alpha$. It  is clear that the Lagrangian is  gauge invariant.

We can think of the states as represented by wave functionals
$\Psi[ \phi, A_{\mu}  ] $. When the potential has the typical
Mexican hat form $V= \lambda (\phi^{*} \phi -v^2)^2$ is when we
expect to have  a spontaneously broken symmetry situation, i.e. to
have a ground state which does not share the symmetries of the
theory. However in this case, this is not only an incorrect choice
of the ground state but is also a choice that is not allowed by
the constraints of the theory. This is particularly transparent in
the Hamiltonian formulation of the above theory. Upon the 3+1
decomposition the action becomes
 \be
S=\int dt \int d^3x \ \vec E \dot{ \vec A }+\pi \dot \phi-{\cal
H}- A_0({\vec{\nabla}}\cdot \vec E+ig (\phi\ \pi-\phi^*\pi^*)) \ee
where $\pi$ and the electric field $\vec E$ are the momenta conjugate
to $\phi$ and $\vec A$ and the Hamiltonian density $\cal H$
takes
the form
\[{\cal H}=\frac{1}{2}\left(\vec E^2+\vec B^2
+\pi\pi^*+(\vec{\nabla} \phi+ig\vec A\phi)\cdot(\vec{\nabla}
\phi^*-ig\vec A\phi^*)\right)+V(\phi\phi^*),\]
 with $ \vec B =  \nabla \times  \vec A$. The zero component
of the vector potential $A_0$ is a Lagrange multiplier associated
with the Gauss constraint:
 \be{\vec{\nabla}}\cdot \vec E+ig(\phi\
\pi-\phi^*\pi^*)=0\ee

It is easy to check that a classical solution minimizing the total
energy of the system is given  by $\vec A=\nabla \alpha, \vec E=0,
\pi=0$ and $\phi=\exp(ig\alpha)\phi_0$ for $\phi_0=$constant
corresponding to the minimum of the Mexican hat potential
$V(\phi\phi^*)$, and $\alpha(\vec x,t)$ and $A_0(\vec x,t)$
arbitrary space-time fields. The fields $A_0(\vec x,t)$ and
$\alpha(\vec x,t)$ are pure gauge degrees of freedom: they are not
observable and their dynamics is undetermined. They have no
physical reality, since they cannot be measured by any physical
system constructed out of $U(1)$ gauge invariant interactions.
This is so unless the gauge symmetry is truly broken, which means
that the full Lagrangian of nature breaks explicitly $U(1)$ gauge
symmetry\footnote{This is not the case in the Higgs mechanism used
in the construction of the standard model.}. Therefore, the
distinction between different choices of $A_0(\vec x,t)$ and
$\alpha(\vec x,t)$ only exist in the notebook of the physicist but
have otherwise no objective meaning. Mathematically, for any two
different choices of fields $A_0(\vec x,t)$ and $\alpha(\vec x,t)$
we do not have different solutions to the theory above but one and
the same solution written in different `field coordinates' or
gauges.

This is why the solution written above does not break the gauge
symmetry in any way. Only if we think of $\alpha$ as a physical
quantity then we can be  mislead to the conclusion (unfortunately
found in many basic textbooks on the subject) that a certain point
on the bottom of the Mexican hat potential is selected by the
classical solution. Customarily, one represents the classical
solution written above in a particular gauge (e.g. $\alpha(t,\vec
x)=0$); however, the ``breaking of the symmetry'' is just in the
choice of gauge and has nothing to do with any dynamical
consideration (such as the form of the potential). Notice that the
situation is different from the global symmetries case already at
the classical level. Even classically, there is no physical
meaning to saying that there is a classical solution sitting
`somewhere' around the $S^1$ defining the minimum of the
potential. Physically, each point around the $S^1$ (gauge orbit)
is to be identified as one and the same state!

The situation in the quantum theory is exactly the same as in the
classical one. This is particularly transparent in the Dirac
canonical quantization of the gauge theory. Let us illustrate this
in the case of our electromagnetic example.  What follows is
rather formal as in conventional treatments one starts by fixing
the gauge in some way which eliminates the question by
construction from the starting point\footnote{ A manifestly gauge invariant
treatment is in principle possible in terms of some non
perturbative treatment such as that of lattice gauge theory.}.

In the Dirac program one defines the so-called auxiliary Hilbert
space $\sH$ of wave functionals of the chosen configuration
variable (e.g. $\Psi[A,\phi,\phi^*]\in \sH$). The canonical
variables also  become self adjoint operators: in the polarization
chosen here $A$ and $\phi$ act by multiplication, while the
electric field $E=-i\hbar \delta/\delta A$ and $\pi=-i\hbar
\delta/\delta \phi$. The (first class) constraints of the theory
are promoted to self adjoint operators which become the generators
of infinitesimal gauge transformations in the quantum theory.
Finally, physical states are required to be annihilated by the
(first class) constraints. So in our case one requires

\be \left[\int_{\R^3} i \hbar {\vec{\nabla}} \alpha \cdot \frac
{\delta}{\delta\vec A} - g\hbar \alpha (\phi\  \frac
{\delta}{\delta \phi}-\phi^* \frac {\delta}{\delta\phi^*})\right]
\Psi[A,\phi,\phi^*] =0, \ \ \ \forall  \alpha(x),
\ee
where
$\alpha(x)$  ranges over a complete set are suitable test fields.

The previous expression can be written in a more familiar form. It
just requires that
 \be \Psi[\vec A,\phi,\phi^*] =\Psi[\vec
A-{\vec{\nabla}} \alpha,(1+i\alpha)\phi,(1-i\alpha)\phi^*], \ \ \  \forall  \alpha
 \ee
which is just the condition that the state be gauge invariant.
Therefore, physical states cannot break the gauge symmetry of the
theory independently of whatever the form of the interaction
potential appearing in the Lagrangian. This conclusion is general
as long as the Lagrangian that defines the theory contains gauge
symmetries. In particular, the state minimizing the energy of the
system must satisfy the above condition which means that it cannot
correspond  to the field being ``localized" anywhere on the bottom
of the Mexican hat potential. A more appropriate image is that of
a wave function whose amplitude is exactly homogeneously
distributed around the hat.

We can proceed further by changing the coordinates in field space
(the space of functions ${\cal F }$  on which the wave functional
is defined). In the discussion above ${\cal F }$   was taken as the
space of triplets $(\vec A,\phi,\phi^*)$  where $\vec A$ is a
smooth vector field on $R^3$ and $ \phi$ is a smooth complex
scalar field on $R^3$.  We will  re-parametrize  ${\cal F }$  so
that now each point is represented by $(\vec A,f ,\theta)$  where
$f= |\phi |$  and $\theta=arg(\phi)$. We must of course be mindful
of the multiple parametrization associated with the
 change $\theta \to \theta +2\pi$.
{Now, we can parametrize the physical configuration space in terms
of the gauge invariant fields\footnote{This is known as a
Stuckelberg transformation.} \be \vec C(x)=\vec A(x)-\vec \nabla
\theta(x) \ \ \ { \rm and} \ \ \ f(x) \ee

In accordance with the discussion above the functional
representing any state of the system must satisfy: \be\label{ggcc}
\Psi[\vec C, f ,\theta] = \Psi[\vec C, f , \theta +\alpha
]=\Phi[\vec C, f ], \qquad \forall \alpha \ee i.e. physical states
are independent of $\theta$ which is a pure gauge degree of
freedom.  In fact in terms of the new variables the Gauss
constraint simply becomes $\pi_{\theta}=0$, where $\pi_{\theta}$
is the momentum conjugate to $\theta$.
  The Hamiltonian density acting on
such states takes the form:
 \ba && {\cal H} \ \Psi[\vec C, f
,\theta]=\nonumber \\ && \nonumber
=\frac{1}{2}\left(-\hbar^2\frac{\delta^2}{\delta \vec
C^2}+(\vec\nabla\times \vec C)^2 - \hbar^2 \frac{\delta^2}{\delta
f^2}-\frac{\hbar^2}{f} \frac{\delta^2}{\delta \theta^2}+
(\vec{\nabla}f)^2+ f^2 \vec C\cdot \vec C+V(f^2)\right)\Psi[\vec
C, f ,\theta]\\ && \nonumber
=\frac{1}{2}\left(-\hbar^2\frac{\delta^2}{\delta \vec
C^2}+(\vec\nabla\times \vec C)^2 - \hbar^2 \frac{\delta^2}{\delta
f^2}+ (\vec{\nabla}f)^2+ f^2 \vec C\cdot \vec
C+V(f^2)\right)\Phi[\vec C, f],
\ea
 where in the last line we have
used the fact that the states satisfy the Gauss constraint (eq.
(\ref{ggcc})).  Note that the term $f^2 \vec C\cdot \vec C$ will be
responsible for the appearance of an effective mass for the vector
field $\vec C$ in perturbation theory around the minimal energy
state for a mexican hat potential $V(f^2)$. More precisely, we can
write \[V(f^2)=V(v^2)+ (1/2) V^{\prime\prime}(v^2) (f-v)^2+{\cal
O}[(f-v)^3],\] where $v$ is the value of the field $f$ at the
minimum of the potential. Then one can  treat the  terms ${\cal
O}[(f-v)^3]$ as (higher order) self-interactions in the
perturbation theory. If we do so the ground state at lowest order
perturbations around the minimum of $V(f)$ becomes
 \be \Psi[\vec C, f ,\theta]_0 ={\cal N} \exp[-
(4\hbar^2) \int C^a(x) [(\triangle + v^2)
\delta_{ab}-\partial_a\partial_b]^{-1} C^b(x)- (f
-v)[\triangle-V^{\prime\prime}(v)]^{-1}(f-v) dx ,\label{vacu}
 \ee
The Gaussian profile with spread $\sigma=v/(2\hbar)$
for the dependence of the wave function for the vector field $\vec
C$ implies the massive character of the latter with an acquired
mass $m=v/\hbar$.  Notice that for the state
 (\ref{vacu}) $ \langle\phi
(x) \rangle =\int [ {\cal D} A_i][ {\cal D} f ][{\cal D} \theta] f
e^{ \theta} |\Psi[\vec A, f ,\theta]_0|^2 =0$ as must be the case
for any gauge invariant, and thus physical state. However it is
quite clear that the expectation value $ \langle\phi
(x)\bar\phi(x) \rangle =\int [ {\cal D} A_i][ {\cal D} f ][{\cal
D} \theta] f^2 |\Psi[\vec A, f ,\theta]_0|^2 \approx v^2$ which
gives rise to the mass of $\vec C$.

\subsubsection*{Fermion masses}

In the standard model  fermions are also thought to acquire mass
as a result of the SSB  and the asymmetric vacuum expectation
value of the Higgs field.  The key difference with the case of the
bosons is that these  masses turn out to be proportional to $v$
rather than to $v^2$. There is a very simple way to describe the
mass generation \cite{Strocchi} in a fully gauge invariant way.
Assume that the fermion fields are $\psi_L, \psi_R$ and that they
are  characterized with the Higgs-electromagnetic Lagrangian
(\ref{hem}) trough:
 \be
  {\sL}_{f}=i\bar\psi_L D_{\mu}
\gamma^{\mu}\psi_L+i\bar\psi_R\partial_{\mu}
\gamma^{\mu}\psi_R+\lambda \phi \bar\psi_L \psi_R 
\ee
 so that under a
gauge transformation $\psi_R$ is invariant and $\psi_L\rightarrow
\exp{i\alpha}\psi_L$. One can define new (manifestly) gauge
invariant fermions $\Psi_R=\psi_R$ and $\Psi_L=\phi^*
\psi_L/(\sqrt{\phi \phi^*})$ (note that the normalization in
the previous definition is needed in order to have the correct
fermion dimension). In terms of manifestly gauge invariant fields
the Yukawa coupling becomes \be {\sL}_{yuk}=\lambda
(\sqrt{\phi\phi^*}) \Psi_L\Psi_R=\lambda f \Psi_L\Psi_R. \ee Due
to the fact that $\langle 0 | (\sqrt{\phi\phi^*})  | 0 \rangle=
v$ the previous term generates a fermion mass for a manifestly
symmetric vacuum.

 We should point out that even if a potential does not have the typical  ``Mexican hat" shape, 
  but rather, say, has  a  simple parabolic profile,  the vacuum expectation value of $\phi ^2$  can
   not be zero (as that would require an infinitely sharp value for $\phi$  which would imply an infinite 
   uncertainty of the conjugate momenta of that filed) and thus  the possibility of obtaining the 
   standard SSB phenomenology exists, in principle,  even   with such  simple potentials\footnote{The detailed 
   estimates for the corresponding values   that emerge in any straight forward analysis,   do 
   not seem to work in the Standard Model of Particle Physics.}.

\section{The Standard Lore}\label{lore}

In  the usual accounts  of the paradigm of SSB in field theories
one encounters an emphasis on the difference between ordinary
quantum mechanical systems of finite number of degrees of freedom
($QM$) and quantum mechanical systems with infinitely many degrees
of freedom ($QM_{\infty}$) such as those treated in the
thermodynamical limit of statistical mechanics and field theory.

The arguments one finds in the literature on the subject,
regarding  the  central difference between  the two situations,
rely on  the existence, in the latter case of unitary
in-equivalent representations of the algebra of observables
\cite{In-equivalent}. This is a well known feature of quantum
field theory, most conspicuously found in the context of quantum
field theory in curved space-times \cite{Wald}.  The situation
arises when the comparison between the states involved indicates
that in going from one to the other necessitates the
application---in the mean\footnote{By ``in the mean''   here refer to  the  difference in the expectation 
value of the total particle number operator between the two  states.
 }---of an infinite number of creation/annihilation operators.
One is  thus, in general, lead to take the so called  ``algebraic
approach" in which all the states in all representation are
collectively considered in a unified fashion, and in practice one
relies on physical considerations and Fell's
theorem\footnote{Fell's theorem states, loosely speaking, that for
any  representation of the canonical commutation relations,  and
for any  set of possible outcomes of
 measurements, one can find a state that  could be  associated with
 these results  to  any desired,  but finite,  accuracy.
  See  for instance \cite{Wald}  page  81.} to
justify working with any one of the Hilbert space representations
that are associated with a particular problem. Nevertheless it is
clear  from these same analyses that  there could exist  in
principle  physical processes mapping states from one irreducible
representation to another.  This will  be  the case,  when the
process  involves the  creation or annihilation of an  ``infinite
number of particles", as it  occurs,  for instance, in the
contexts of cosmological evolution, or  systems in  interaction
with certain ``external potentials". 

   The fact that one needs to deal with physical processes involving
 transition from one to the other of the in-equivalent constructions,
 or folia, is what lead to the development of the so called Algebraic
 Approach, which allows for the simultaneous treatment of all states
 in all follia (see for instance \cite{Wald}). We should be aware
 that even in describing one single simple physical situation from two
 different points of view, requires one to pass from one QFT
 construction to an in-equivalent one, as exemplified by the analysis
 of the Unruh effect \cite {Unruh}.
  Thus, it seems clear that, in
 general, physics might not force us to describe the state of the
 system, as belonging at all times, to one of these representations.
 In fact, if that was the case, we could not consider the phenomena of
 phase transitions associated with a spontaneously breaking of the
 symmetry, as transitions occurring in physical time, which is
 precisely what one often wants to do. These
 considerations indicate that the central question we need to confront
 is: What are the physical, rather than mathematical, issues behind
 the ``phenomena of spontaneous symmetry breaking".  Stating that we
 face ``in-equivalent representations of the algebra of observables"
 does nothing to indicate the physical reason for the breaking or not of a symmetry present in
 the laws governing the relevant phenomena by the lowest energy states
 of the system in question.  We must understand when does this
 happen, if at all, and in those cases, the physical reasons behind
 it.

We note that in those same accounts  of the issues
such
in-equivalence of representations are often invoked as the
fundamental difference between $QM$ and $QM_{\infty}$ which makes
possible the phenomena of SSB in the latter but not in the
former \cite{Strocchi}.  Those analyses  start by tying this rather
mathematical issue (the  in-equivalence of representations) to
various, apparently  sacrosanct physical characteristics, namely:
i) the ``clustering property" and the ii) ``irreducibility of the
representation" (\cite{Strocchi} page 119).

Let us review these points here briefly to uncover what
assumptions are really behind the standard picture. The starting
point is to consider the algebra of observables generated
(by algebraic plus Cauchy completion)
by the local observables $A_{loc} $\footnote{This algebra  is
constructed by  integrals of the local field operators over
compact space-time regions.}. Such algebra satisfies, by
construction, the property of asymptotic locality\footnote{This
property,  sometimes called ``Asymptotic Abelianess",  states that given two local observables $\hat A\& \hat B$.
if we define by $A_x$ the displaced observable $ e^{i \hat P x}\hat Ae^{-i \hat P x} $  then
 in the limit $x\to \infty$ the observables commute i.e. $ lim_{x\to \infty} [A_x, B] =0$.}, justified by the realization that in
a laboratory, only those states that are associated with local
operations can ever be created or measured (pages 34, 142 and
156 of \cite{Strocchi}). Then, one uses the result ( page 35 of \cite{Strocchi}) that
warrantees that,  given an
 algebra satisfying the asymptotical locality conditions, and
assuming that one has a cyclic\footnote{A state is  said to be
{\it cyclic},  if  when  acted upon by  all the elements of the
algebra in question, it  leads to a dense set on the Hilbert
space.} translational invariant vacuum,  and if moreover  the representation
is irreducible, then  there can be no other translational invariant
state. In the case of relativistic fields, the cluster property is
strongly tied to the uniqueness of the ground state, by the Arakki-
Hepp-Ruelle Theorem,  (see for instance page 65 of \cite{Strocchi}).

The issue we need to face,  is then:  the physical  relevance
of the following  properties:
\begin{enumerate}

\item Asymptotic locality:

As we indicated, the assumption of validity of the  property of
asymptotic locality is generally  well justified in the usual applications, by the realization that,
in a laboratory, only those states that are associated with local
operations can ever be created or measured.  The problem is, of
course, that when we want to concern ourselves with issues that
involve processes that are relevant for the evolution of our
universe, such limitations to local operations have no clear
justification.  The issue of whether in our universe the vacuum is
invariant or not (assuming that the true theory is invariant
under a certain continuous global symmetry with the   conditions
normally associated with SSB), must then be answered on different
grounds.

In particular, the notion of  asymptotic locality looses its mathematical
meaning in the framework of a spatially compact universe (as the example
treated in Section \ref{Cont}). We have explicitly shown in that example that
the ground state is  indeed symmetric, that the symmetry is unitarily implemented, and
have argued that the standard local physics can be described in such framework. Of course,
loosing asymptotic locality might take us out of the hypothesis of some of the
strong theorems of local quantum field theory. One of the standard results that is no longer
valid is clustering decomposition.

\item The clustering property:

 This property corresponds to the the condition  where  the two point  field
correlation  for  the vacuum  state  decays to zero at long
distances. The requirement of  this  property is usually  justified in physical terms
that are often misleading. The demand  that local experiments
should ``not depend" of what happens at space-like separated regions, taken in unqualified terms,
is a severe  and unphysical restriction in quantum mechanical context:
we know that
 non-local correlations do occur in situations such as an EPR experiment, and
that they have indeed been observed experimentally \cite{Aspect}. Moreover,
contrary to what is often loosely stated, and in close analogy with what
occurs in the EPR correlations, the violation of the clustering property (in
the sense of the existence of correlations, wether or not they are associated
with the vacuum state) does not need to lead to conflict with relativistic
causality. The statement should  then be replaced by a more limited
requirement that information should not propagate
superluminally\footnote{Consideration of this issue will necessarily bring
into consideration the question of measurement in quantum field theory, an
issue which has difficulties of its own\cite{Sorkin}.} and these two different
requirements should not be confused.
  What is clear
from the analysis of such situations is that such correlations
cannot be used to send a-causal signals by any observer to another
\footnote{In fact  there seems to be an unjustified  time-bias in
our description of physics regarding the possibility of
correlations.  We naturally  expect them  to  arise  after  an
interaction but {\it a priori} dismiss the possibility of  
they  exiting before the interaction\cite{Hu Price}.}.

From the perspective of the Wightman axioms, the clustering
property is a theorem rather than a requirement. All the same, the
restrictive conditions that lead to this result are hidden in the
axioms themselves.  Of crucial importance is the requirement of
the uniqueness of the Poincar\'e invariant state directly tied to
the irreducibility of the representation (namely, property (2)
above).

We have seen that the spatially compact model of Section \ref{Cont}
violates the previous two properties as a consequence of the quantum nature
of the zero mode and that this leads to no problems at all. It should thus be clear that  the
 violation of  such properties  in the $ L\to  \infty$ limit   can not lead to any  contradiction 
 either, and that the consideration of the symmetric vacuum in this case is not prevented by
 any physical argument. A particularly clear exhibition of a characteristic of such state is the infinite range
field correlations in the symmetric vacuum shown  in equation (\ref{indy}).

\item Irreducibility of the representation:

This requirement is linked to the assumption (one of Wightman
axioms \cite{Wightman}) that the state space of a relativistic
quantum field theory is a separable Hilbert space. Despite of the
fact that separability of the Hilbert space is a powerful
mathematical tool leading to a variety of important theorems
applicable in a vast variety of physical situations, it is clear
that this does not include all physical
situations\footnote{Perhaps  the best known example  of a physical
theory where one needs to deal with a non-separable Hilbert space,
is the case of  Loop Quantum Gravity \cite{LQG}.}.

Contrary to the previous two properties, the present one is not
 violated by the spatially compact model analyzed in Section
 \ref{Cont}. However, this property is often violated in physically
 relevant situations (e.g. thermodynamics).

\end{enumerate}

In order to illustrate the issues mentioned above in a concrete and simple
context, let us consider the one dimensional infinite spin chain model
(infinite Ising model). It is the theory of an infinite number of spins placed
along the real line and interacting with the closest neighbors. The possible
states of the system are labeled by the infinite series of elements
$\{(\alpha_i,\beta_i)\}|_{i\in \Z}\in
\H_{\infty}=\cdots\C^2\otimes\C^2\otimes\cdots$, modulo an overall phase. The
Hilbert space $\H_{\infty}$ is a simple example of von Neumann's infinite
tensor product Hilbert space \cite{vonnewman}. In the Dirac notation a state
$|\Psi\r\in\H_{\infty}$ is written 
\be 
|\Psi \rangle= \cdots \otimes |
s_{-1}\rangle \otimes | s_{0}\rangle\otimes |s_{1}\rangle\cdots\otimes |
s_{j}\rangle\otimes\cdots 
\ee
 where $|s_i\rangle$ is a state of a spin half
particle.  The inner product between two such states is simply given by the
infinite product of the corresponding spin half states, in the appropriate
limit, of course: 
\be
 \langle \Psi |\Psi ' \rangle= \lim_{ N\to \infty }
\prod_{ i =-N}^N \langle s_i | s'_i\rangle\label{vonn}.
 \ee This inner product
is well defined because it should be clear that unless two states differ in
just a finite set of places the inner product will vanishes, and if they only
differ in finitely many places it reduces is a finite product.  It is clear
also that $\H_{\infty}$ is ``very large", indeed it is not even separable. For
instance if we denote $ \lbrace |\vec n,+\r ,|\vec n,-\r \rbrace$ the basis of
the Hilbert space of single particle spin states associated with the direction
$\vec n$, we can define translational invariant states of $\H_{\infty}$ as
 \be
|\Psi^{\pm}(\vec n)\r=\cdots \otimes |\vec n,\pm\r\otimes |\vec n, \pm
\r\cdots,\label{ejemplito}\ee by which it is understood that all the spins in
the infinite chain are in the same state $|\vec n,\pm \r$. Then one has
$\l\Psi^+(\vec n)|\Psi^+(\vec m)\r=0$ unless $\vec m=\vec n$. Moreover, for a
given direction $\vec n$ one also has $\l\Psi^+(\vec n)|\Psi^-(\vec n)\r=0$,
as it follows immediately from (\ref{vonn}). In fact any local excitation of
$\Psi^{+}(\vec n)$, where by local excitation we mean a modification of the
state on a finite number of sites in the infinite chain, will remain
orthogonal to any local excitation of $\Psi^-(\vec n)$ or $\Psi^{\pm}(\vec m)$
respectively. One can therefore separate $\H_{\infty}$ into orthogonal worlds,
each of which provides an irreducible representation of the algebra of local
observables.  These are separable Hilbert spaces $\H_{\vec n,+}$ ( and
$\H_{\vec n,-} =\H_{-\vec n,+}$ ), constructed as follows: one first chooses a
state $\Psi^{+}(\vec n)$, and constructs $\H_{\vec n,+}$ by the Cauchy
completion of the set of states obtained from $\Psi^{\pm}(\vec n)$ by the
application of the complete collection of operators that act on finitely many
of the spins (those belonging to the algebra of local observables).  In each
sector one can see that the corresponding translational invariant state, which
will be called ``the vacuum of that sector" is by construction a cyclic
state. Moreover, the representations $\H_{\vec n,+}$ for different $ \vec n $
are unitarily in-equivalent\footnote{The in-equivalence one refers to, in
this context, is the fact that the operators realizing the mapping from one
sector of the full Hilbert space to the other, have vanishing matrix element
when restricted to states in each sector.  The different sector are on the
other hand ``unitarily equivalent" in the sense of the existence, among the
sectors, of a biyective mapping which preserves the inner product.}. All these
are characteristic features of $QM_{\infty}$.

If we consider the direct sum $\H_{\vec n,+}\oplus\H_{\vec m,+}$ this gives a
reducible representation of the algebra of local operators, and in that case
there are more than one translational invariant vacua (it is customary to
refer to generic states of this type, as ``mixed phase states"). Indeed the
combination $\Psi=\alpha\Psi^{+}(\vec n)+\beta \Psi^{+}(\vec m)$ with $
|\alpha |^2 + |\beta |^2 = 1$, are similarly translational invariant (but
clustering is violated due to long range correlations). The full Hilbert space
$\H_{\infty}$ is a highly reducible representation of the algebra of local
observables with uncountably many translational invariant vacua. The reason that
these reducible representations are somewhat exotic in the standard QFT
applications is that they do not satisfy the famous clustering property;
essential among other things for the definition of scattering theory. As
indicated before this property has been overvalued, and assertions that
without it we could not do physics cannot be sustained (recall EPR discussion
above).

At first sight, and in the mindset of Fell's theorem, there seems to be no way
to distinguish through finitely many local measurements wether we are in a
pure or mixed phase. In the above example it is clear that no matter what
local quantity one considers, it is always possible to obtain the same
expectation values, probabilities, etc. by taking a mixed phase
$\Psi=\alpha\Psi^{+}(\vec n)+\beta \Psi^{+}(\vec m)$ or a suitable state in a
pure phase (one just needs to tune the state in the pure phase to get the
right answers). Does this mean that breaking or not breaking the symmetry
becomes a meaningless statement? We will argue that the answer to this
question is, in general, in the negative. We must start by recognizing that in
our practice of physics, we  do often consider things that are not strictly
related to what we measure or will measure\footnote{For instance we might
consider the behavior of certain quantity, as we measure it at different
scales, and if the result of all previous measurements indicate that such
quantity does not change, we might consider extrapolating, and as a working
hypothesis consider that they never change. Strictly speaking the
corresponding measurements would be possible, only trough nonlocal
experiments, and it seems clear that the ``limit when the distances between
the observations regions tend to infinity" can not, in practice, be carried
out by humans. Nevertheless, we can envision experiments carried out at very
large distances whereby one could ``reasonably infer" the behavior of the
correlations as the distances between the points involved increase without
bounds, and these inferences can be taken to give meaning to the notion of
``correlations between points infinitely separated".  This way of reasonably
accessing in practice, these sort of limits, are reminiscent of the issue of
``practical measurements" of probabilities, which, in the frequency
interpretation, are defined in terms of infinite number of trials. It is clear
that we never make experiments involving infinite trials but, nevertheless, we
do consider, as measurable, certain probabilities which are defined based on
such idealizations. It is clear then, that we do often set different criteria
of ``measurability" in different fields of physics, and we must, then, be very
careful when considering the corresponding intersections}. In the situation at
hand, we find, for instance, that certain infinitely long range correlations
(clustering violation) are present in the mixed phase and we might want to
consider such state as the state appropriate for describing the extrapolation
beyond the actual measurements (which, in practice, can only concern finite
distances), of a situation we might conceivable find by local experiments.
More concretely, in the field of theoretical cosmology we often consider
situations based on the cosmological principle, and it would be inappropriate
then to shackle oneself to employ a quantum field theory description in which
such principle can not be incorporated.  More generally, if we have a
situation in which the internal symmetries of the field theory can only be
realized for certain type of states, one can not argue that the symmetry must
be broken, simply because one has {\it a priori} limited (by construction)
the range of states under consideration based on such ``measurability
arguments".

In the simple setting of the present example it is clear that even when many
of the usual questions can be addressed using the irreducible representations
$\H_{\vec n,\pm}$, the full physics in the one dimensional spin chain is
contained in $\H_{\infty}$
\footnote{I.e.  there is no reason to {\it a priori} exclude from
  consideration as a possible description of the world, any of the states in
  $\H_{\infty}$}.  In fact some of the justifications often invoked in the
  literature to try to justify a consideration of one sub-sector of the
  theory, in these sort of situation can be turned around.  For instance, the
  very same argument showing that, once in a pure phase $\H_{\vec n,\pm}$,
  there exist no local process that can take us away from it, serves to show
  that if we take the system to start off in a mixed phase $\H_{\vec
  n,\pm}\oplus\H_{\vec m,\pm}$ there is no local process, which will allow the
  system to evolve into a pure phase (elements of say $\H_{\vec n,+}$).

Let us return to the question of symmetries in the context of our simple
infinite spin chain model.  Let us focus for simplicity on the discrete
symmetry $z \rightarrow -z$.  The ``mixed state" $|{\rm\bf
mixed}\r=\Psi^+(\hat z)+\Psi^-(\hat z)$ is clearly symmetric under that
discrete symmetry.  However, there exist other translational invariant states
respecting that symmetry. In particular, let us consider the state in which
each individual spin in the chain is in the state $| \hat x,+ \rangle =
\frac{1}{\sqrt 2 } ( |\hat z,+\rangle +|\hat z, - \rangle $). This can be
taken as a ``vacuum state" $|{\rm\bf pure} \r=\Psi^+(\hat x)$ and is clear
that it is symmetric under $z \rightarrow -z$. It corresponds to a ``pure
phase".  Its is worthwhile warning the reader against confusing the two states
$ |{\rm\bf mixed}\r $ and $ |{\rm\bf pure}\r$, or thinking that a
translational invariant symmetric state in a theory like the one considered in
the present example, must necessarily, be a state of the form $|{\rm\bf
mixed}\r$.

Having said all this we must also point out that in a theory like
the one under consideration, all the discussion above, does not
have any bearing on deciding which is the true vacuum: a state
that breaks the symmetry, such as $\Psi^+(\hat z)$,  one that
does not  but respects   the clustering property, such as $
|{\rm\bf pure} \rangle $, or one like $|{\rm\bf mixed}\r$ which is
symmetric but breaks the clustering property. It is clear that
either of the first two can be used as a reference state to lie at
the basis of the perturbative construction of a Hilbert space.
Either can be taken as the unique translationally invariant in
the physically relevant representation.    Regarding the    third,
we believe that there are no  convincing  physical arguments
indicating that it  should not  be considered as a possibility.
 Of  course, the one   which  is  appropriate to a specific
situation  will depend in general on the details of that
situation.

\section{  Further Considerations: Energetics and the thermodynamical limit}

Having analyzed in detail the quantum field theoretical
constructions most commonly associated  with SSB,  we believe that
is  appropriate  at this point  to  take a more global view of the
relevant issues and discuss  some aspects  that often lead to
misinterpretations  and to  inappropriate ``reasonings by analogy".
The following discussion  is motivated, in large part, by our own
state of confusion before embarking on this  analysis,  and it is
our hope that it  will  serve our colleagues who have  been equally
puzzled by the ``standard lore" on the subject.

The usual  discussions on the present subject start by stating that, in the case of a quantum mechanical
system with a  finite number of degrees of freedom, the vacuum state is
unique\footnote{As we have seen that requires in particular
 that the Hilbert space  does contain symmetric states.
 We have encountered  already simple cases  where this does not occur.
    }   (i.e. non-degenerate) and  possesses the symmetries  of the Hamiltonian.
That situation is often then contrasted with what occurs when the system has an  infinite number of d.o.f..
The way this contrast is  often  presented,  starts by considering a  quantum  mechanical
particle  in a double well potential. It is then argued that  if
the system is prepared  to be localized in one  of the wells  it
would  tend to tunnel into the other well  with a nonzero
probability  thus indicating that it was not in the ground state,
and that such state would indeed be symmetric. This basic
picture  is  not  believed to cover systems with an  infinite
number of degrees of freedom  like those contemplated in the
thermodynamical limit in statistical mechanics or in  field
theories due to the infinite energy barrier that separates  the
different minima of the energy functional. This difference is
often  invoked to argue that in a system with an infinite number
of degrees of freedom the ground state need not share the
symmetries of the theory and   thus that  the symmetry could  be
spontaneously broken.

We find this argument misleading as it mixes up two issues that
must be considered separately.  One is the issue of how long would
a system that is prepared in a non-symmetric state stay in that
state, and a  very different  one concerning  which  is the  lowest energy
state.   It might well be that the system once prepared into a
non-symmetric state  would never evolve towards the full symmetric
state, or even that it might stay in the initial state forever,
but that does not mean that the initial state was the true ground
state.

Let us  start by considering  a system made of infinite number of
particles  each of which is in a double well potential. The
 naive ``energetics" argument  can be used to argue  that if  the system is  initially
prepared in a state with all the particles wave functions picked on
the left well will be quasi-stationary (i.e. would take an infinite
amount of time to tunnel to a state in which all the particles are
in the right well),  something that  is obviously wrong.  However, one might in fact  compute the
energy of the barrier and argue that an infinite time of tunneling  should be expected in association
with the fact that this barrier is infinite. However, in this case,
we can treat each of the noninteracting degrees of freedom
separately and evaluate the wave function of each of them at any
finite time after the preparation time. On the other hand  we note  that each particle is  initially in a state
described by a wave function that has support only on  the left, and as the
different degrees of freedom are not interacting, the subsequent  evolution of each particles' wave function,
cannot differ from the corresponding  evolution  extracted from the evolution of the whole system and thus we
conclude that at a finite time after the preparation of the  full system,
the  state will not be the initial one, and thus, that it could not
possibly have been the ground state. We see in this trivial example
that the standard arguments based on heuristic energy considerations
cannot be taken at face value.

Of course the above example lacks something that is common to the
systems with infinite number of degrees of freedom that are
usually  considered in this regard:  An interaction that tends to
drive all the degrees of freedom of the system towards the same
individual state for that degree of freedom d.o.f.. This is the
case for instance for  a ferromagnetic sample of infinite extent,
where the interaction amongst the neighboring  magnetic moments
tends to align them. This feature then implies that  there can be
no transitions associated to an  individual d.o.f.\footnote{We are
assuming that  there are no sources of energy capable of providing
the energy represented by a single  d.o.f. becoming misaligned  with the rest.}, but the
allowed transitions involve a coordinated and simultaneous change
in all the system. Thus it is the interaction amongst all the
individual subsystems that stabilizes the asymmetric state of the
global system. This however does not indicate that the system was
in its ground state.

There can exist, in principle,  stable states other than the true ground state,
and all that is needed for such  stability is the absence,  for some reason, of
perturbations connecting these states with others of lower energy.
The existence of an infinite energy barrier  in systems known to
have a  finite energy is  the best known example of such
situation.  In the case of systems that are infinite in extent
however, the issue of finiteness of the systems energy is
something that cannot be treated  lightly, and more careful analysis is required to
reach  reliable conclusions  one way or the other.   In fact the
thermodynamic limit involves taking the number of particles  $N\to\infty$ ( often stated
as $V\to\infty$ limit while keeping the value of the intensive
quantities  (like pressure  $P$ or  temperature$T$) fixed  awhile  those of the
extensive quantities scale with $N$ so $E/N$ or $E/V$ (where $E$ is the system's energy and $V$ its volume) is kept
constant.  In those situations the total energy of the infinite
system is infinite and  the arguments involving  infinite energy
barriers  cannot be taken  as reliable.

\subsection{Thermodynamical limit and a system in equilibrium
 with a thermal reservoir}

One of the  paradigmatic  examples often used to illustrate  the
phenomena of  SSB occurs in the thermodynamical treatment of
various systems. The lessons  extracted are then translated
essentially unchanged to the field theoretical situations. We would
like to clarify the similarities and differences between the two
situations in order to uncover some misleading analogies often found
in the literature. In order to do this it is convenient to review
the setup that lies behind the statistical mechanical treatment as
it involves often forgotten assumptions that are not always valid in
the quantum field theoretical context.

The starting point of such treatment is  to consider  the  system of
interest  as one  within an ensemble  of  identical systems (as far
as  their dynamics  is concerned) and then  study the behavior of
the ensemble averages of the quantities  of interest. Under certain
physical conditions (which are often, explicitly stated, in defining the  statistical mechanical
systems), the  ensemble
averages are expected to coincide with the time averages of the
corresponding quantities in the particular system (i.e. element of the ensemble).
The point however is that important new assumptions go into the
theoretical description the quantum mechanical statistical mechanical
system (among which the large, or even infinite number of degrees of
freedom is only one). To start with, we must keep in mind the
fundamental role of the interaction of the system with external
d.o.f.. At the practical level, this is characterized by the
impossibility of complete isolation of the system, and in particular
by the need interact with it  to carry out   observations, in order to draw any conclusions. At the
theoretical level, the many external d.o.f. with which the system of
interest interacts are represented in the statistical treatments by
a thermal bath. Notice that even in the case where we set $T=0$, the
external environment has a fundamental influence, being the physical
source of the de-coherence that is brought into the analysis in the
form of the hypothesis  of {\it a priori} random phases which is
needed to justify the construction the micro-canonical, canonical,
and grand canonical ensembles in the statistical mechanics of
quantum systems (see \cite{Pathria} page 109, Ch5.2 A).
Therefore, one key feature of thermodynamical systems is their
intrinsically open nature.

 As mentioned above, one of the basic
assumptions of statistical mechanics is the ergodic hypothesis: that
is the notion that the average over the members of the
ensemble---thought to be spread over the accessible
micro-states---coincides with the time average for an ensemble
member over the curse of its dynamical evolution. In addition to the
assumption of equilibrium, the ergodic hypothesis relies, generically,  on the
existence of a large enough set of uncontrollable external
disturbances that affect the system in minute but essentially random
fashion. These disturbances prevent the system from remaining in a
relatively small set of states during the course of its evolution
and are though to  imply ergodicity (see \cite{O. Penrose}).

The intrinsically open nature of thermodynamical systems is a
crucial difference with SSB situation in quantum field theory
studied in Section \ref{Cont}. This justifies the description of
system ``in equilibrium" in interaction with a thermal reservoir at
fixed temperature $T$ by the Gibbs density matrix $\rho(T)$ (instead
of pure states) given by\footnote{There is a precise analogy between
statistical (thermodynamical) systems and certain quantum field
theory situations  even when the latter describe quantum
mechanically closed systems. This analogy comes from the familiar
Euclidean formulation of QFT where expectation values (in the
appropriate vacuum state) of observables can be computed by the
formula \be <\phi(x_1)\cdots\phi(x_n)>=\int D\phi
\phi(x_1)\cdots\phi(x_n) \exp{(-S_{eucl}[\phi])}, \label{40}\ee where
for simplicity we have chosen  as an example a scalar field theory. The
assumptions implying the validity of the previous equation
(justifying the validity of Wick rotations) are unrelated to those
implying the validity of (\ref{free}).},
 \be \rho(T)
=Z^{-1}e^{-\frac{H}{kT}}=Z^{-1} \sum_{i}e^{-\frac{E_i}{kT}} \
|\psi_i\rangle\langle\psi_i|, \label{free}
\ee
 where
$\{|\psi_i\rangle \}$ is the set of the energy eigen-basis with
eigenvalues $E_i$. The previous equation will
be important for discussion of SSB in the ferromagnetic system. The
latter is often used as a paradigmatic example of SSB in general;
however, as we argue in the following section the analogy is
misleading.

\subsection{ The ferromagnetic phenomena.}

One situation that is often used as  an illustrative example of
the phenomenon of spontaneous symmetry breaking is the spontaneous
magnetization of ferromagnetic materials. In this section we
review this example and show in what sense symmetry is broken by
the ferromagnet below critical temperature. The model often
considered is given by a system with large number $N$ of spin half
atoms arranged in a lattice where the translational degrees of
freedom are taken as frozen. Other possible internal degrees of
freedom of the individual atoms are ignored. Each atom is assumed
to have a ferromagnetic interaction with its nearest neighbors. If
one labels the spin of the $a$-th atom by $\vec s_a$, the
Hamiltonian of the system in interaction with an external magnetic
field $\vec B$ is:
 \be
 H = -J \sum_{a,\ b=1}^{N} \vec s_a  \cdot \vec s_b  - \mu \vec B\cdot \sum_{a=1}^{N} \vec{
 s_a}.
 \ee
 where $J$ and $¼mu$ are coupling constants. It is customary to simplify the analysis of the  system and
concentrate instead on the Issing model. The system is analyzed in
terms  of the canonical ensemble at fixed temperature $T$.  In this
situation one considers the Helmholtz free energy $A(T,B)$, and one
obtains the magnetization as $M(B,T) =- (\frac{\partial A}{\partial
B})_{|T} $.   The spontaneous magnetization corresponds to the
situation in which $M(0,T)\not=0$ which occurs in general for
$T<T_c$ for a certain critical temperature $T_c$. This  type of
analysis  is in general in excellent accord with the experimentally
observed features of ferromagnetic materials, such as  details of
the phase transitions and the appearance of magnetization domains,
etc.

Let us consider the issue of symmetry breaking. What symmetry is
broken when spontaneous magnetization occurs? The first important
aspect that should be emphasized (and which is a main difference with the
examples considered in Section \ref{Cont}) is that the Hilbert
space of each of the individual d.o.f. does not posses symmetric
states. More precisely, suppose we start with the single spin state $|\vec
n,+\rangle$---in the context of the spin chain example of Section
\ref{lore} and in the same notation---one could try to construct a
rotational invariant single particle state by superimposing spin
states in all possible directions. This could be concretely
achieved by acting on $|\vec n,+\rangle$ with a Wigner rotation
matrix ${\rm \bf D}^{1/2}(g)$ for $g\in SU(2)$ and then group
averaging by integrating the resulting state on all possible
rotations with the invariant measure (the $SU(2)$ Haar measure in
this case). The invariant state would look like 
\be
|{\rm \bf
invariant}\rangle\equiv\int_{SU(2)} dg\  {\rm \bf
D}^{\frac{1}{2}}(g)|\vec n,+\rangle.\label{st}
\ee 
However, is easy
to see that the previous average vanishes identically, namely
$|{\rm \bf invariant}\rangle=0$. Moreover, one can easily show
that the absence of a non-trivial rotationally symmetric state
also holds for a system with finite number of spins. Thus there
cannot be a rotationally symmetric state in the full Hilbert space
(with infinitely many spins) simply because there is no symmetric
state in the Hilbert space of the individual d.o.f. or any finite
number of d.o.f..

This establishes an important difference between  a ferromagnet (or
the spin chain of Section \ref{lore}), which exhibits phenomena such
as the spontaneous magnetization, and the scalar field theories of
Section \ref{Cont} with their characteristic `` Nambu-Goldstone boson"
phenomena, or the gauge theories of  Section \ref{gshf} where  one
finds the Higgs phenomenon.

Thus the question remains: if there are no quantum mechanical symmetric states
in what sense symmetry is broken by the ferromagnet below critical
temperature? The key point is the statistical mechanical nature of the
treatment of the system. Indeed the ferromagnet at temperature $T$ represents
an open quantum mechanical system in equilibrium with an external environment,
and is therefore described by a density matrix of the form (\ref{free})
instead of a pure state. Now, despite of the fact that no quantum (pure) state
of the ferromagnet can be rotationally invariant, the density matrix can
indeed be symmetric. For example, at $T=0$ the sphere-worth (degenerate)
ground states of the ferromagnet dominate (\ref{free}) and the resulting
density matrix is rotationally invariant. For instance the density matrix
$\rho_{inv}^{(1)}$ for a single spin \be\rho_{inv}\equiv \int_{SU(2)} dg\ {\rm
\bf D}^{\frac{1}{2}}(g)|\vec n,+\rangle ({\rm \bf D}^{\frac{1}{2}}(g)|\vec
n,+\rangle)^{\dagger}=\frac{1}{2}\ \mathbbm{1}.\label{den}\ee For $N$ spins
one gets $\rho_{inv}^{(N)}=P_{N/2}/(N+1)$ where $P_{N/2}$ is the projection
operator into the irreducible representation space of spin $N/2$ contained in
the tensor product space of $N$ spin half particles, as expected. Thus the
density matrix describing the low temperature magnet is spherically symmetric
yet the possible states in which the magnet can be found when an element of
the ensemble is singled out by observation breaks the rotational symmetry by
selecting one of the sphere-worth of degenerate vacua\footnote{ Here we must
be careful and avoid confusing two situations that are often mixed up.  The
density matrix can be used to represent the state of an ensemble of identical
systems, where each individual element of the ensemble is in a pure state.
One can use the statistical matrix to represent a system that has no
individual state of its own, as in the case where the system of interest is
part of a larger system which, as a whole, is in a state involving correlations
between the subsystems in such a way that the subsystem of interest can only
be described trough its reduced density matrix.  Finally we could have an
ensemble of subsystems as in the previous case. }.  This is the sense in which
rotational symmetry is broken by the ferromagnet below critical temperature.

Therefore, two important facts lead to a notion SSB in the case of
the ferromagnet: On the one hand the statistical mechanical
treatment of the system; which is valid under the conditions that
justify the assumptions of ergodicity, and {\it a priori}
uncorrelated phases. On the other hand, the absence of symmetric
pure states in the quantum theory which lies at the heart of the
phenomenon. These features are not shared by many other systems,
often compared with the example of the ferromagnet, as for instance
those that are usually presented in discussions of the emergence of
a Goldstone mode and in the associated discussions of  the Higgs
phenomena.

\subsection{Two Kinds of  Phase Transitions}

We  should  be  aware  that there are in fact  two notions of
phase transitions  that despite having several  features in common
are   conceptually  quite different  and it  is thus important  to
keep those differences in mind  when considering analogies.  The
thermodynamic  phase transitions   which have been studied
extensively  and  the  purely quantum phase transitions  that have
received  much less  attention (see however  \cite{ZurekQPT,
Sachdev}). The latter  are often  characterized as ``changes in
the character the ground state of the system and are  described
via unitary reversible dynamics". Thus  they are thought to  be
quite different from the ordinary thermodynamical phase
transitions that are  in general  associated with an irreversible
process. To think of a concrete example  we might consider the
quantum Ising model corresponding to a set of collection of spins
subject to an interaction among nearest neighbors  corresponding
and also to an external magnetic field  with a Hamiltonian of  the
following form:
    \be H = - g B  \sum_{i} \sigma_i^x  - J\sum_{< i,j>} \sigma_i^z\sigma_j^z
    \ee
where $g$ and $J$   are  a coupling constants  $B$ is external
magnetic field  along  a fixed  direction  $x$,  the  $\sigma$s
are  the standard Pauli spin matrices  and the  second  sum is
over near neighbor pairs.  We  note that in this model the spins
are thought to interact only trough their  $z$ components  while
the external field  acts in a perpendicular direction.    Thus
while  the internal interaction tends to align  the $z$
components of the various spins,  the  external  magnetic  field
tends to align  the $x$  components of all spins  in the direction
of the field.

We first concentrate in the case  where  the  system  is not in
contact  with any  other external degree of  freedom,   and thus
can be treated  through  the exact Schroedinger equation for an
isolated system  as   there is no channel through which  it can
loose  quantum phase coherence.  We  can  consider the changes in
the nature of the ground state of the system and see  that  by
changing the magnitude of the field  we can bring about a change
in the character of the ground state. When $B$  is  sufficiently
large the ground state  corresponds to a situation in which all
spins  are align in either the $+x$  or $-x$ direction (depending
on the sign of $B$),  while  in the  case   where $B$ vanishes,
there is  in fact a  two dimensional   space  of   ground states
with  a  basis given by the two degenerate  states   corresponding
to  all spins  pointing in the $+z$  direction,  and all spins
pointing in the $-z$ direction. This  degeneracy  is ignored in
the discussion of \cite{ Sachdev},  probably  because  the focus
there  is on different issues.   As  we now consider   the changes
in the nature of the manifold of ground states  as $B$ is
continuously changed  we encounter   a clear  example  of  a
quantum phase  transition.  What  we must note here,  however, is
that for  all  values of $B$,  the manifold of   ground states
includes  states that  are invariant under reversal of the $z$
axis,  exactly  as we found  in  the discussion of section I C.

Let us now consider the same system when in contact with a thermal reservoir
at a finite temperature $T$. We can now consider the standard statistical
mechanical treatment of such situation.  We know that if we consider changing
this temperature (for a fixed value of $B$) we encounter the usual
thermodynamic phase transitions.

The first difference we encounter among these two types of transitions, is that
when studying a system in contact with a thermal reservoir, the most stable
state is not the state of minimal energy but a state with minimal free
energy. These free energy minimizing states can be thought to be found among
the possible states of the system by a suitable optimization process in which
besides the obvious changes which lower the energy and increase the entropy,
one might include changes that involve increasing the systems energy if they
are accompanied by a sufficiently large increase in the entropy.  The fact
that the lowest energy state of symmetric theories corresponds to the most
symmetric state (or at least includes such a state in the set degenerate
vacuua) might, in these situations, be superseded---as far as the
determination of most stable state is concerned---by a large amount of entropy
that is associated with the possible complicated correlations of the systems
degrees of freedom with the larger set of less symmetric states for the
d.o.f. of the environment.

One might be tempted to think that in the $T=0$ limit the statistical
mechanical treatment would always coincide with the quantum mechanical of
quantum field theoretical treatments, and that thermodynamic phase transitions
go smoothly into quantum phase transitions.  That this is not necessarily the
case has been observed before, for instance in the discussion is section 1.2
of \cite{ Sachdev}.  We should point however that the consideration of the
issues of symmetry forces us to look with additional light on this matter.
The point is that in the case where the vacuum state is degenerate, the
statistical mechanical analysis, by its reliance on representative ensembles,
assumes that all the vacuum are equally weighted in the description of the
system, while a quantum mechanical treatment would normally not involve such
assumption and could consider one individual system as described by any of the
various degenerate vacua.  In fact the statistical treatment in such situation
does not allow within its formalism (i.e. in the partition function
description) the consideration of issues such as which basis should be used to
characterize the various manifold of vacua sates of the individual system (i.e
turning to the example of section I C should the natural description be in
terms of non-symmetric states, symmetric pure states or symmetric mixed
states?). By default, one describes the whole ensemble with a density matrix
which would be symmetric even in the cases where there are no symmetric states
as discussed in the previous subsection. There is however no way to even
consider the issue of what is the state of one of the individual elements of
the ensemble.

One of the points where it is clear that the statistical mechanical treatment
can hide some of the issues we have been confronting in this paper lies in the
construction of the micro-canonical statistical ensembles for quantum
mechanical systems. There one brings in an {\it ad-hoc} assumption about the
relative phases of the states of the various systems in the ensemble called
the hypothesis of {\it random a-priori phases} (See section 5.2 A of
\cite{Pathria}). This hypothesis is akin to imposing de-coherence in the
particular and, in principle, arbitrarily preselected basis, when one turns to
apply the formalism to describe the state of an individual system (i.e. to
describe the state of each of the elements of the ensemble when there are
degeneracies, and in particular for the situation where the ground is
degenerated and we want to consider the $T\to 0 $ limit).

Thus  the set of differences between the situations described
within quantum field theory for an  isolated system  and those
described with the  quantum theory of an open  system   are
sometimes  rather  stark,  and   we must keep  them in mind  if we
want to be able to trace to the appropriate  source  of some
behavior  we  are trying to understand. It might  well be that in
most circumstances, the statistical mechanical treatment  of open
systems is the appropriate one, and in fact most of the  examples
of  observed   phase  transitions are described within that
paradigm.  Among the few  experimentally observed exceptions of an
example of a truly  quantum phase transitions exists (see
\cite{Sachdev}).

The    above considerations  and  in particular the caveats in the
analogies  and  the  several distinctions  between the two sets of
phase transitions and their treatment,  show that  it would be
clearly erroneous to  consider that  the well known examples of
thermodynamic  phase transitions  represent,  in any sense,  some
sort of   incontrovertible evidence about the correctness of  all
of the  standard lore  about the  subject  of   SSB in  quantum
filed theories.

\section{ The Cosmological Setting}

This  should  be,  in some sense, the paradigmatic object of our
discussion,  as the universe  is perhaps  the  only truly isolated
quantum system\footnote{Even the most isolated laboratory system
can not be decoupled from the gravitational d.o.f.  and  the
extent of such coupling  has been  argued  to be enough to force
one to consider the  system as open \cite{Kay}.}. However, as we
have already mentioned, this  setting is in fact rather
 the inappropriate  one to  consider the question of  what is the true
vacuum of the theory,  as  the  fact that  in this  situation we
do not have at our disposal  a well defined  notion of energy
(because the  general situation is  not stationary),   the
question of  which is  the state that minimizes the energy,
becomes, strictly speaking, meaningless. Moreover, if  any definition  of global energy of the system would become available
it  seems  clear the states that are compatible with large scale
homogeneity and isotropy (i.e. the cosmological principle)   would
all have infinite energy unless we were dealing with a  finite
universe.

One possibility  that one has to have in mind in discussing these
issues in the cosmological context is that the
universe might be closed, and thus, all the issues associated with
the infinite extent of the region where two states differ,  which
we saw are fundamental for some of the conclusions,  and the
arguments based on  boundary conditions which lie at the basis of
the analysis of  Section II C, would loose their  relevance.  In fact
it seems natural to think that the issue of the universes large
scale topology  should have no bearing on the conclusions.

Nonetheless we face in this context,  and in particular in the
inflationary scenarios,  one of the earliest examples of what one
would like to  call  the spontaneous  breaking of  a symmetry. The
point is that inflation is supposed to take  the universe (or at
least  the relevant part of it) to a state that is totally
homogeneous and isotropic, in both  its   classical background
(the inflating FRW with the slow rolling inflaton field)  and  its
quantum  fluctuations (of course  we view this as an approximated
description of a full  quantum state of the system, which is
however thought  to  share the symmetries of its  simplified
description).  The usual account for the origin  of the initial
spectrum of fluctuations   sees them  as emerging from the quantum
uncertainties of an homogeneous and isotropic  vacuum state.  We
have  argued elsewhere \cite{Ours-Inflation}  that the standard
accounts of this process fall short of being fully satisfactory
and that additional  elements must  be called upon, beyond those
usually considered in this context. The present analysis  should
serve to remove the  standard account of  SSB as part of the
answer to the conundrums raised in \cite{Ours-Inflation}.

When  dealing with cosmology,  on the other hand, one might need
to face the question of whether our universe should be considered
infinite or finite, and one might be tempted to wonder wether
these type of issues would  offer  a possible avenue to
empirically address such questions. Unfortunately the fact  is
that realistic considerations would require not only the
addressing  of a full set of unsettled  cosmological  issues, but
the need  to  be in possession of a well defined notion of energy,
which is generically  not available  in the situation at hand due
to the lack of time translation invariance of the space-time and
the impossibility to assign an energy to the gravitational degrees
of freedom. Therefore it is clear that nothing of that sort can be
envisaged at this time.

Another related issue is what can we take to be the initial conditions of our
universe in regard to the symmetries in question (those we would have wanted
to associate with the zero mode), i.e. were the initial conditions symmetric
or not?. This takes us to the realm of quantum cosmology, where one can hope
to find some relatively convincing information about the initial state in one
of various proposals \cite{Quantum Cosmology}, or alternatively to the realm of
inflationary cosmology \cite{Inflationary Cosmology} in which one expects the
universe to eventually reach a stage where the conditions become independent
of such initial data, at least in a large enough region containing all of our
causal past. It seems clear that in such situations, even if one can simply
take the universe to be spatially infinite, the question of which is the
appropriate ``vacuum" has changed dramatically as compared to the other
situations treated in this manuscript, as one can no longer rely on simple
energy considerations, and then the question of wether the state of the system
is or not symmetric, must be addressed on different grounds. In this sense,
there does not seem to be any argument that would justify the conclusion that
the "vacuum" must break the symmetry of the theory. We will briefly touch on
some related issues in the context of our analysis of topological defects.

   \subsection{Topological defects}
     One of the issues that is closely tied with the standard accounts of SSB
      is that of the generation of topological defects.  In the case where the
      vacuum is degenerated, the idea that the theory (or more correctly, the
      system under the appropriate conditions) ``selects" locally one vacuum,
      among the various possible ones, naturally leads to the possibility that
      the vacua selected in various spatial regions are not only distinct but
      that they are, collectively, ``knotted" in such a way that the resulting
      configuration becomes stable due to a large energy barrier that
      separates it from the, un-knotted global minima of the energy.  These
      ``knotted" configurations are called topological defects an are supposed
      to be the inescapable consequence of the dynamics of the early universe.
      Similar considerations can be applied to some condensed matter systems
      which lead to analogous topological defects in the context of some
      thermodynamical phase transitions (see for instance
      \cite{Pnematic-Fluids}).

  The treatment of these issues is carried out in the classical language that,
  we have seen, can be very misleading in dealing with the situation in
  quantum field theory.  In particular we have seen in section II, that the
  true vacuum, for the restriction of the system to any local region (which
  has finite spatial extent), is unique and is symmetric, whenever the theory's
  Lagrangian is symmetric.  
  system is therefore locally allowed to relax to the vacuum, the different
  regions will be brought to the same state, and no knotting can be
  expected\footnote{The remaining issue is whether there is some mechanism
  that would create the correlations between the local states, and which are
  needed if these local states are to come together and form a global
  vacuum \cite{WaldCorrelations}.}.
  This
  raises the question: Do we, in such cases expect the formation of
  topological defects?  We will focus here on the case of continuous global
  symmetries and leave the case of Gauge Theories for future analysis.  Let us
  consider, for concreteness the case of a theory with an $O(4)$ symmetry in a
  3+1 static homogeneous and isotropic space-time background with closed
  spatial slices (Einstein's static universe).  In the cosmological setting we
  would be interested in considering an expanding universe, but for the sake
  of our discussion we consider here the static example in order to have a
  well defined energy associated with every field configuration. Let $t$ be
  the Killing ``time" parameter and consider the foliation of this space-time
  by the 3 spheres of constant $t$ : $\Sigma _t$.

  Now let us consider the usual account of the generation of
  topological defects.  To do so we imagine the field configuration to
  be created a time $t_0$.  For the sake of our discussion, we imagine
  that the $\Sigma_{t_0}$ is divided in $N$ patches, the configuration
  is created independently in each patch and then some interpolating
  smoothing is carried out on a region of width $D$ in the
  intersection around of the patches.  The first thing to note is
  that, as in the quantum setting the vacuum state is unique, in
  contrast with what happens in the classical account, if in each
  patch the state reproduces the vacuum locally, then it
 will reproduce the vacuum globally\footnote{At this point
  one could become concerned with the issue of relative phases among
  the local states,  but this is an issue that is different from
  the one we are dealing with here.  In fact the phases have group structure
  $U(1)$ independently of the group structure of the symmetry which is
  supposed to undergo SSB.}. Thus in this case one should not expect
  the generation of any topological defect.  But let us assume for a
  moment that for some reason, the local states reproduce, not the
  vacuum but one of those states in which the field is sharply peaked
  around one of the classical vacuum, a state similar to  that of eq. \ref{noway}. Let us
  refer to such sate as ``centered on the classical vacuum pointing in
  the direction $\phi^{(0)} _1$" and denote them by $|\phi^{(0)} _1
  \rangle $.  Let us imagine, in fact, that the global state is indeed
  close to a classical configuration with nontrivial  topological
  winding number.  The issue is the topological stability of the
  quantum configuration.  The point is that, while classically the
  field configuration space is a mapping $\phi: S^3 \to R^4$ which has
  a degenerate  set of minima, with topology $S^3$, and thus the configurations
  that are restricted to be locally at the minimum correspond to maps
  from $S^3$ to $S^3$ which we know are labeled by  the winding number\footnote{This particular type of topological defect is called a
  {\it texture} \cite{Texture}.}, at the quantum level what we have is
  a state, characterized by a wave function: a mapping from
  $\Psi:F(S^3, R^4) \to C$ where $F(S^3)$ is the set of functions 
  $\phi$ from $S^3$ to $R^4$.  The topological characteristics of the
  two sets of mappings need not be the same.

    Let us consider for instance the temporal evolution  of the state $
     |\phi^{(0)} _1 \rangle$. The state is homogeneous and thus should
     have no component in the mode $k$.  Such state corresponds to a
     wave packet of eigen-states of $\pi_0$, and since this operator
     commutes with the Hamiltonian, each of the components will evolve
     just in its phase, as in the case of a free particle.  It then
     follows that the wave packet would spread in field space (as a
     free particle spreads in position space).
       As this
     happens the wave packet will become less and less concentrated
     about $ \phi^{(0)} _1 $.  When we recall that the components
     corresponding to the various $\pi_0$ eigenvalues remain present
     throughout the evolution we conclude that no matter how large
     this spreading becomes, the evolution will not transform the
     state in the true vacuum, a conclusion that is evident from
     energetic considerations.  For that to happen one would need some
     interaction that leads to dissipative phenomena as far as this
     mode is concerned.  In a more realistic cosmological context the
     expansion of the universe could be one source of effective
     dissipation, although by itself it could not disrupt the charge
     conservation   associated with $\pi_0$.

Having gained this insight we now turn to the situation that is
relevant to the issue of topological defects.  Let us consider first
the case in which we have only two contiguous regions such that in each
one of them the state locally resembles the state $\phi^{(0)} _1$ and
$\phi^{(0)} _2$ respectively.  How would this state evolve?  First of all it
is clear that far  away from the boundary the state should evolve in the
same way as in the preceding discussion, i.e.  the wave packet would
spread and the sharp value of $\phi^{(0)} _1$ and $\phi^{(0)} _2$
would become less and less sharp.  Now we note that in this case we
are not really dealing with the zero modes as the spatial dependency
of the state indicates the strong excitement of the modes with
wavelengths $\lambda$ of the order of the size of the two regions.

These modes are not the ones associated with charge conservation and so there
is nothing to prevent the energy to flow into other modes.  This de-excitation
of the modes can be expected to continue until the energy is equally spread
between all modes not protected by charge conservation.  This suggests that the
evolution of the state will consist of a simultaneous spreading of the wave
packet, and de-excitation of the $k \sim \lambda^{-1}$ modes leading to an
erasing of any ``memory" of the values of $\phi^{(0)} _1$ and $\phi^{(0)} _2$
.  Thus except for the relative small excitations in all modes, the global
state will evolve into something closely resembling the true vacuum.  If we
add to this any source of dissipation, as that associated for instance with
the cosmological expansion, we can expect to approximate the true vacuum to be
approaches to an increasing degree as the system evolves.

      This view is of course in sharp contrast with that inspired by the
      classical picture.  Particularly striking is the contrast in expectation
      of what is the fate of a state that approximates the classical
      configurations that correspond to topological defects.  The previous
      discussion suggests that such configuration would evolve towards the
      vacuum and that the classical notion of winding number, is made
      irrelevant because at the quantum level such notion is not well defined
      and the number we can associate to a state in some approximate
      semiclassical description, is simply not conserved.

   How can we view the unwinding of a state that approximates a classical
    configuration which corresponds to a topological defect?  An important
    aspect that must be kept in mind to help understanding how that can occur
    lies in the distinction between $\langle V(\phi)\rangle$ and $
    V(\langle\phi\rangle)$.  If we are dealing with a classical state
    corresponding to a static configuration containing a topological defect,
    the unwinding would require the field to pass trough configurations that
    have at some points the value $\phi =0$ and such evolution can be made
    impossible by the large energy density associated with $V(0)$.  In the
    quantum language, on the other hand, the state of the field is described by
    a wave functional, and as such, the relevant quantity in such consideration
    is the behavior of $\langle V(\phi)\rangle$ as the state of the system
    evolves. As it is clear from the example provided by the true vacuum state
    in sub-section II A, this quantity can be very different from $V(0)$, even
    when the expectation value for the filed is zero. Thus the state of the
    system can evolve in such a way that the expectation value of the local
    field $\langle\phi (x) \rangle$ can go trough zero in some points --and
    even to zero globally, if there are some sources of dissipation--
    unwinding in the process the ``topological defect" without the impediment
    posed by the large energy barrier associated with $V(0)$.

The  study of the actual evolution of  a state which approximates
a classical configuration corresponding to  topological defect is
a  very interesting problem, which we hope to approach with
suitable numerical techniques in  future works. Given the expected
complexity of such analysis  its study  lies clearly outside the
scope of the present article.  Nevertheless it is easy to present
a toy evolution that exhibits  the possibility of  a quantum
unwinding of  a  configuration that is classically ``knotted", and
thus impossible to unwind classically. Let $\Phi[\phi(x)]_{def}$
be a wave function for the filed $\phi(x)$   which corresponds to
a classical topological defect, and  let $\Phi[\phi(x)]_0$. Now
consider the wave function $\Phi[\phi(x)] (t) = \alpha (t)
\Phi[\phi(x)]_{def} + \beta (t) \Phi[\phi(x)]_0 $ where $\alpha
(t) $  is a function  such that $\alpha (0)=1  $ and $\alpha
(\infty) =0$  and
   $\beta  (t) $  is a function  such that $\beta  (0)=0  $ and $\beta (\infty) =1$   and  are  normalized to preserve the
    norm of the sate throughout the evolution.     This mock evolution,
     while completely unrealistic, illustrates  the important differences between the quantum and classical pictures
     that stands behind the distinct  fates of the topological defects as can be expected from
     the two kinds of  analysis.  Needless is to say that the physically relevant description, is the quantum version
     because we currently view the classical description  of any system as nothing more than a useful approximation valid only
     under  certain conditions  and as long as  its results  do not conflict  with the  corresponding quantum analysis.

\subsection{ Topological defects in Statistical Mechanics}

  Topological defects are known to arise as a result of phase transitions in
  many systems.  There is a widespread notion that this is a common feature of
  quantum phase transitions and statistical mechanical phase transitions.
  However we must recall that in the case of the former, the issue one must
  address in considering the equilibrium state of the system is not just
  related to the minimization of the energy, but with the tendency of entropy
  to increase, and thus one must consider minimizing the Helmholtz free energy,
  rather than the energy of a pure quantum state.  Under such circumstances
  the precise correlations that are exhibited by the true vacuum of the
  isolated quantum mechanical system, are statistically unlikely and as such,
  they would be associated with a very low value of the entropy.  One can
  think of this as due to the essential random nature of the interaction of
  the systems d.o.f.  with the ``environment".

The point is that the states that minimize that kind of quantity are in
general not pure quantum mechanical states but rather mixed states, which are
characterized, among other things by a large and complicated set of
correlations between the system of interest and the collection of systems
which play the role of thermal reservoir or ``environment". Under those
conditions one can envisage a situation where individual regions that arrive
at local equilibrium and that are thus described by a thermodynamical state
--{\it and the point is that in this case there are in fact many different
such states}--that has locally minimized the free energy, end up in a global
configuration such that each local state would differ from those corresponding
to the other regions in such a way that the global state would correspond to a
topological ``knot".  This is in fact what lies behind the appearance of
textures, domain walls, strings and monopoles, in suitable systems when they
undergo thermodynamical phase transitions of different kinds.  The point
however is that this situation is quite different from the one we have been
interested on throughout this manuscript and which concerns the case of
quantum mechanical isolated systems.

\section{Discussion}

 In the present manuscript we have considered the type of situations where SSB
  is usually thought to arise.  We have seen that the claim that such
  situations are characterized  by a vacuum state that does no share the
  symmetries of the theory, are often incorrect.  

In particular in the field theoretical settings, where the quantum system has
  an infinite number of d.o.f., and is spacially compact with no boundaries
  (box with periodic boundary conditions) the true vacuum of the theory is
  unique and symmetric. This remains true in the case where the spatial extent
  of the system is infinite while lowest energy states become highly
  degenerate including both symmetric and asymmetric states.  We have seen
  this by considering in detail an example showing that the quantization of
  the mode associated with the phase or orientation of the vector in internal
  space corresponding to the bottom of the potential is like that of the so
  called zero mode of a mass-less scalar field.  In contrast with the other
  modes, that are quantized like harmonic oscillators, this mode must be
  quantized like a free particle.  In the case of interest considered in
  section III the field corresponds to a particle on a circle (i.e. a bounded
  region) and so its ground state corresponds to a constant (i.e. a wave
  function that is independent of the value of the field) where the field has
  no defined value.

In the case of quantum field theory on a spacially non compact space, the
symmetry breaking nature of the customary notions of vaccum is a feature of
the mathematical model of QFT and not a physical phenomenon. The question of
whether the vacuum state is symmetric is simply ruled out as the zero mode of
the field is not one of the dynamical degrees of freedom in local quantum
field theories. The symmetry issue has infrared subtelties in QFT. If one
wants to properly talk about the symmetry of the vacuum in QFT one needs to 
bring in a infrared regulating structure that would allow one to consider the
zero mode as a fully dynamical degree of freedom. The simplest way to do this
is to define QFT in a box with periodic boundary conditions.

 We have shown that the standard phenomena usually associated with SSB like
   the emergence of Nambu-Goldstone bosons, the Higgs mechanism and Fermion
   mass generation, and the analysis of approximate symmetries such as the
   Chiral Symmetry of the QCD with light quarks, can all be recovered in a
   treatment where the vacuum state is symmetric. { That is, that all the
   successful phenomenology usually attributed to SSB is fully recovered
   if we define QFT in an infrared regulating box (with periodic
   boundary conditions) for a sufficiently large box, and thus all physical conclusions remain
   unaltered and can be understood within the context of the analysis
   presented throughout this manuscript in which the main feature is the
   complete symmetry of the vacuum in symmetric theories}. 

Possible  observational consequences of our analysis are not ruled out.
One example is the prediction of long range correlations
(clustering violation) independent of the infrared regulator Equation (\ref{indy}), or
the suggestion (Section \ref{aps}) that the uncertainties in the vacuum of certain
operators in QCD might be large due to the approximate axial 
symmetry of the ground state.


  We have argued, on the other hand, that the standard account for generation
  of topological defects in connection with the SSB in the field theoretical
  context is rather suspect, and that the classical topological stability of
  the resulting objects, might not be extensible to a quantum treatment.

  We have further considered briefly a non-isolated quantum mechanical system
    and its possible states, together with the set of hypothesis that underly
    the statistical mechanics considerations, which are in general not
    applicable to the case of an isolated quantum system.  We have indicated
    that in those circumstances something akin to a spontaneous symmetry
    breaking might occur when considering the most stable states, which
    should not be confused with the true ground state of the system. One might
    well argue that the real world examples are better represented by the
    former and that the latter is an unjustified idealization.  This might
    indeed be right for the vast majority of cases of interest, but even then,
    it is worthwhile understanding where does the justification for the
    treatment lie, i.e. in the fact that the system is not isolated? or in the
    characteristics of ground state of the isolated system's Hamiltonian?
    Moreover, keeping these issues in mind is fundamental to uncover
    situations where the usual treatment might not be justified, and where
    other considerations might have to be brought to bear in understanding the
    behavior of the system in question.  Such is the case where one is
    considering issues like the spontaneous symmetry breaking of quantum field
    theories which one would hope characterize the complete set of existing
    fields, and where no obvious external d.o.f. can be called upon to play
    the role of thermal environment, or aspects of quantum cosmology where the
    system under consideration is the whole universe and thus one is dealing,
    by definition, with an isolated quantum mechanical system.  It might be
    that one can, in some of those situations, find a different way to justify
    the standard treatments, but it is important not to deceive oneself into
    believing that the standard justifications apply automatically to those
    cases as well.

{ We end by emphasizing that the view we have presented here, does not only
clarify the issues but liberates us from the very uncomfortable position of
having to rely on the infinitude of our universe as part of the justification
of the standard accounts for the possibility for the phenomena of SSB that one
finds in the more careful accounts of these issues \cite{Symmetry Breaking}.}

\section*{Acknowledgments}

\noindent
 We thank Marc Knecht for very useful discussions, and also Chryssomalis
 Chryssomalakos and Alberto Guijosa for reading the manuscript and pointing
 some problems we corrected.  Needless is to say that any remaining errors
 would be ours and only ours.  It is a pleasure to thank the support of
 L'Universite de la Mediterranee for an extended visit by DS to the Centre de
 Physique Theorique, in Marseille.  This work was supported in part by
 DGAPA-UNAM project IN119808.


\begin{thebibliography}{99}


\bibitem{In-equivalent}
  R.~Haag,
  ``Local quantum physics: Fields, particles, algebras'',
{\it   356 p. (Texts and
monographs in physics, Berlin, Germany: Springer  (1992))}

\bibitem{Wald}  R. M. Wald, {\it``Quantum Field
Theory in Curved Sapacetime
and Black Hole Thermodynamics"},
(University of Chicago Press, 1996).


 \bibitem{Unruh} W.G. Unruh  ``Notes on black hole evaporation"
 {\it  Phys. Rev.  D }{\bf 14}, 870, (1976).


\bibitem{Aspect} ``Experimental realization of Einstein-Podolsky-Rosen-Bohm Gedankenexperiment:
A New violation of Bell's inequalities"
 A. Aspect, P. Grangier, G. Roger,
{\it Phys. Rev. Lett.} {\bf 49}, 91, (1982).


\bibitem{Wightman} 
  R.~F.~Streater and A.~S.~Wightman,
  ``PCT, spin and statistics, and all that,''
{\it  Redwood City, USA: Addison-Wesley (1989) 207 p. (Advanced
book classics)}.

\bibitem{Nair}  V.~P.~Nair,
  ``Quantum field theory: A modern perspective'',
{\it  New York, USA: Springer (2005) p 557 }.


\bibitem{Elitzur} ``Impossibility of spontaneously breaking of a local symmetry",  S Eliutzur,  {\it Phys. Rev. D} {\bf 12 }, 3978, (1975).

\bibitem{EPR}  See for instance discussions about  the  EPR experiment in  A. Peres
``Quantum Theory: Concepts and Methods" (Kluwer Academic  Publishers, 1993)

\bibitem{Mermin}
 ``Is the Moon There when nobody Looks?"
 D. Mermin {{\it Physics  Today }} {\bf 32}, 38, (1985).

\bibitem{WaldCorrelations} ``Correlations beyond the horizon"
R.M. Wald, {\it Gen. Rel. Grav.} {\bf 24},1111, (1992).

\bibitem{Pathria} ``Statistical Mechanics", R. K. Pathria, (Butterworth-Heinemann 1997).

\bibitem{Rjaraman} ``Solitons and Instantons", R. Rajaraman, (North Holland Publishing  Co. 1982).

\bibitem{Strocchi}
  F.~Strocchi,
  ``Symmetry breaking,''
{\it  Berlin, Germany: Springer (2005) 203 p}; F.~Strocchi,
  ``Elements Of Quantum Mechanics Of Infinite Systems,''
{\it  Singapore, Singapore: World Scientific (1985) 179 P. 
(International School For Advanced Studies Lecture Series, 3)}.


\bibitem{Symmetry Breaking} ``Symmetry Breaking", {\it Lecture Notes in Physics} , F. Strocchi ( Spinger,  Berlin Heidelberg 2005).

\bibitem{Coleman}   ``There are no Goldstone Bosons in two dimensions", S. Coleman,
{\it Commun. Math. Phys.} {\bf 31},  259, (1973).

\bibitem{NO 2 D SFT}  B.  Schroer {\it Fortschr. der Physik }{\bf 11}, 1, (1963). For a more accesible reference  See  for instance  footnote in page 525  of  C. Itzykson and  C. B. Zuber, Quantum Field Theory (Mc Graw Hill 19080). 

\bibitem{ Standard Model}  Among the recent and up to date  accounts  of  the ``Standard Model" of particle physics,  see  The standard Model A Primer,  C. Burgess and G. Moore (Cambridge University Press, U.K.,  2007).

\bibitem{LQG} See for example, ``Loop quantum gravity'', C.
 Rovelli,
 {\it Living Rev. Rel.} {\bf 1}, 1 (1998)
 [arXiv:gr-qc/9710008];
 ``Quantum geometry and gravity: Recent advances'', A. Ashtekar,
[ arXiv: gr-qc/0112038];
 ``Introduction to modern canonical quantum general relativity'', T. Thiemann,
 [arXiv: gr-qc/0110034].

\bibitem{Sorkin} ``Impossible measurements on quantum fields'',  R. Sorkin, in ``Directions in General Relativity"  B. L. Hu, T. Jacobson Eds. ( Cambridge U. Press,
 Cambridge, 1993).

\bibitem{Hu Price} `` Time's Arrow and Archimedes' Point",  Huw Price, (Oxford University Press, UK 1996)

\bibitem{vonnewman} J.~von Neumann,
``On infinite direct products''. Compositio Mathematica, 6 (1939),
p. 1-77.

\bibitem{Jackiw:1995be}
  R.~Jackiw, ``Diverse topics in theoretical and mathematical physics,''
{\it  Singapore, Singapore: World Scientific (1995) 514 p}

\bibitem{Kay}  ``Expectation values, experimental predictions, events and entropy in quantum gravitationally decohered quantum mechanics",
Bernard S. Kay, Varqa Abyaneh,  
 arXiv: quant-ph/ 0710.0992.

\bibitem {Sachdev} S. Sachdev,   ``Quantum Phase transitions", (Cambridge University Press, UK 1999).

 \bibitem{ZurekQPT}  W. H. Zurek, U. Dorner, P. Zoller ``Dynamics  of Quantum Phase Transitions" ( arXiv:cond-mat/0503511v2 ).

\bibitem{Quantum Cosmology} ``Quantum Cosmology  Problems for the 21${}^{st}$ Century", J.~B.~Hartle, [arXiv:
gr-qc/9701022]; ``Generalized Quantum mechanics for Quantum Gravity", J. B. Hartle,
[arXiv: gr-qc/0510126].


\bibitem{Inflationary Cosmology}  ``Quantum Mechanics of the scalar field in the new
  inflationary Universe",
  A. Guth and S.-Y. Pi {\it{ Phys.  Rev.  D}} {\bf 32}, 1899, (1985);
 ``Fluctuations in the Inflationary Universe",
S. W. Hawking {\it{Nucl. Phys.}} {\bf B 224}, 180,  (1983);
``Origin of Structure in the Universe"
J.J. Halliwell and S. W. Hawking,
{\it Phys. Rev. D}, {\bf 31}, 1777,(1985).


\bibitem {Ours-Inflation}
``On the Quantum Origin of the Seeds of Cosmic Structure"
 A. Perez, H. Sahlmann, and
 D. Sudarsky,   {\it Class, and Quant. Grav.}  {\bf 23}, 2317, (2006)
[arXiv: gr-qc/0508100].


\bibitem{Pnematic-Fluids} ``Late time coarsening dynamics in a nematic liquid crystal",
Isaac Chuang, Neil Turok , Bernard Yurke 
{\it  Phys. Rev. Lett. }{\bf 66}, 2472, (1991).

\bibitem{O. Penrose} Penrose O. {\it Foundations of Statistical Mechanics},
(Pergamon Press, Oxford, 1970).

\bibitem{Texture} W. B. Kibble  {\it J. Phys. A}  {\bf 9} 1387 (1976).


\end{thebibliography}
\end{document}